\documentclass[preprint,amsmath,amssymb,superscriptaddress,aps,a4]{revtex4}
\usepackage{xcolor}
\usepackage{graphicx}
\usepackage{amssymb}
\usepackage{multirow}
\usepackage{array}
\usepackage{braket}
\usepackage{diagbox}

\newcommand{\bfk}{{\bf k}_{\perp}}
\newcommand{\bfz}{{\bf z}_{\perp}}
\newcommand{\bfq}{{\bf q}_{\perp}}
\newcommand{\bfb}{{\bf b}_{\perp}}

\newcommand{\bfp}{{\bf p}_{\perp}}

\usepackage{xcolor}
\begin{document}
\title{Role of higher twist distributions in the tomography of proton}

\author{Navpreet Kaur}
\email{knavpreet.hep@gmail.com}
\affiliation{Department of Physics, Dr. B.R. Ambedkar National Institute of Technology, Jalandhar, 144008, India}

\author{Shubham Sharma}
\email{s.sharma.hep@gmail.com}
\affiliation{Laboratory for Advanced Scientific Technologies of Mega-Science Facilities and Experiments, Moscow Institute of Physics and Technology (MIPT), 141700, Dolgoprudny, Russia}
        
\author{Abi Jebarson A}
\email{abijebarson@gmail.com}
\affiliation{Department of Physics, Dr. B.R. Ambedkar National Institute of Technology, Jalandhar, 144008, India}
        
\author{Harleen Dahiya}
\email{dahiyah@nitj.ac.in}
\affiliation{Department of Physics, Dr. B.R. Ambedkar National Institute of Technology, Jalandhar, 144008, India}
	
\date{\today}
\begin{abstract}
We have studied the higher-twist distributions of the proton, including T-even and T-odd transverse momentum-dependent parton distributions (TMDs). Under the umbrella of the light-front framework, we have chosen two distinctive approaches of quark-spectator systems for comparison, one inspired by the soft-wall AdS/QCD and another with a dipolar form factor at the nucleon-quark-diquark vertex. The comprehensive picture at higher-twist provided by both T-even and T-odd TMDs not only aids deeper insights into the internal structure of the proton in the quark sector but also provides an interpretation of different components of the energy-momentum tensor in quantum chromodynamics. Hence, using these standard parton distribution functions, further predictions regarding the physical insights of gravitational TMDs in momentum space are also provided.
\end{abstract}
%
\maketitle
\section{Introduction}
Deep inelastic scattering (DIS) has long served as a cornerstone for studying the complex multidimensional internal structure of hadrons in terms of their partonic constituents \cite{Collins:2004nx, Polchinski:2002jw}. Considering the asymptotic freedom in quantum chromodynamics (QCD) at high energies \cite{Gross:1973id, Politzer:1973fx}, the DIS cross section is approximated by the convolution of lepton-parton scattering with parton distribution functions (PDFs). The parton model, subsequently elevated to the formalism of factorization that provides the theoretical basis for present-day analyses of high-energy hadronic scattering \cite{Ji:2004wu, Collins:1989gx}. 
In the theoretical description of these scattering, the product of two current elements is expanded in the form $J^\mu(y) J^\nu(0)=\sum_{a,l}C^{(a)}_l(y^2)y^{\mu_1} \cdots y^{\mu_l} O^{(a)}_{\mu_1 \cdots \mu_l}(0)$, where $O^{(a)}_{\mu_1 \cdots \mu_l}$ corresponds to the irreducible symmetric traceless tensor of rank $l$. A power counting of mass dimensions gives the singularity structure of the coefficients in the form $C^{(a)}_l(y^2)\sim (y^2)^{-d_J-l/2+d_O^{(a)}(l)/2}$,
with $d_J$ and $d_O^{(a)}(l)$ as the dimension of current $J$ and the local operator, respectively. In the operator product expansion, the scaling of singularities is determined by the twist $d_O^{(a)}(l)-l$ of the local operator $O^{(a)}_{\mu_1 \cdots \mu_l}$ \cite{Gross:1971wn, Alday:2010zy}. The smaller the twist, the stronger the coefficient singularity and the greater the contribution of the operator to physical observables. The coefficient function in the operator product expansion of the DIS hadronic tensor is proportional to the $(M/Q)^{d_O^{(a)}(l)-l-2}$, where $Q$ corresponds to the momentum transfer in the collision and $M$ is the mass of a target in DIS \cite{Qiu:1990xxa, Bacchetta:2006tn, Meissner:2007rx, Meissner:2009ww}. PDFs contain information only about the longitudinal momentum fraction $x$ carried by an active quark from its parent hadron. Further, if we consider the transverse motion of partons, we can get a class of non-perturbative functions, named as transverse momentum-dependent parton distribution functions (TMDs) that provide the information of a parton 
with longitudinal momentum fraction $x$ and transverse momentum $\bfk$ (GeV)
\cite{Diehl:2015uka, Pasquini:2008ax}. Therefore, TMDs comprise the three-dimensional tomography of the partonic distributions in the momentum space. Twist-2 effects in TMD admit a theoretically 
well-established framework and are well supported by experimental data \cite{Burkardt:2008jw, Barone:2010zz, Aidala:2012mv}. Nevertheless, the first measurements of asymmetries in semi-inclusive DIS, involving unpolarized scattering by EMC \cite{EuropeanMuon:1983tsy, EuropeanMuon:1986ulc} and longitudinally polarized targets by HERMES, revealed twist-3 contributions in the fixed-target kinematics of these experiments \cite{HERMES:1999ryv, HERMES:2001hbj, HERMES:2002buj}.

In the past few years, attention has been drawn to higher-twist distributions, which provide finer details at the partonic level.
The structure of T-even higher-twist TMDs for the proton has been explored within different theoretical approaches, including the covariant parton model, the light-front constituent quark model, the diquark spectator model, and the MIT bag model
\cite{ Bastami:2020rxn, Lorce:2014hxa, Pasquini:2018oyz, Sharma:2022ylk, Sharma:2023wha}. Whereas limited work has been conducted in the sector of T-odd TMDs, which requires the higher terms of gauge‑link structure of the quark correlator in terms of initial or final-state interaction (ISI or FSI) \cite{Boer:1997nt, Belitsky:1997zw, Ohnishi:2003mf, Lu:2012gu, Liu:2021ype}. The complete parametrization of T-even as well as T-odd TMDs can be found in Ref. \cite{Sharma:2024lal} for the case of spin-1/2 particles from twist-2 to twist-4. The point-by-point extraction of a twist-3 PDF has also been investigated through the analysis of the beam-spin asymmetry for dihadron production in semi-inclusive DIS off a proton target at CLAS and CLAS12 \cite{Courtoy:2022kca}. In the present paper, we focus on the higher-twist distribution of TMDs for the lightest spin-1/2 baryon, i.e., proton, which is considered as a two-body system, comprised of an active quark (that participates actively with an incoming real or virtual particle) and a spectator diquark in the light-front framework. This diquark can only have the spin 0 (scalar) or spin 1 (axial-vector).
Furthermore, these axial-vector diquarks can be considered as isoscalar and isovector spectator diquarks. In the quark-diquark picture of a proton, ISI/FSI involves an exchange of a gluon between the active quark and the spectator diquark. This interaction can be incorporated either into the light-front wave functions (LFWFs) or via a kernel in the overlap form of LFWFs. We considered a second approach to calculate the T-odd TMDs of higher-twist distributions that modify only the parton's transverse momentum, without altering the quantum numbers of the spectator diquark \cite{Pasquini:2019evu}.
	
In addition to the study of mass, spin, and mechanical properties through generalized parton distribution functions, the connection between TMDs and the QCD energy-momentum tensor (EMT) has recently been computed in Ref. \cite{Lorce:2023zzg} to study the momentum densities, flux of inertia, and stress distribution in momentum space. To do so, the authors have parametrized the EMT for both spin-0 and spin-1/2 in terms of gravitational TMDs and summarize the relations between gravitational TMDs and the standard leading and higher-twist T-even and T-odd TMDs. Their work demonstrates that higher-twist effects are not merely artifacts of power-suppressed contributions but instead encode the response of hadrons to the spatial and mechanical properties of quarks inside hadrons. For a proton target, in general, there are 32 independent functions, and due to the particular structure of the gauge-invariant canonical EMT for quarks, this number reduces to 16, half of which can directly be expressed in terms of the standard quark vector TMDs. 
	
In the present work, we consider two different light-front models to study the higher-twist T-even and T-odd TMDs with the inclusion of a gluon exchange by considering perturbative $U(1)$ gluon scattering kernel. To investigate these partonic distributions, two distinct phenomenological descriptions have been adopted, namely spectator diquark model (SDM) \cite{Bacchetta:2008af} and light-front quark diquark model (LFQDM) \cite{Maji:2016yqo}. One model characterizes the proton-quark-diquark vertex by a dipole form factor. Another model is based on the AdS/QCD correspondence, which provides a holographic framework for connecting QCD dynamics with gravitational dual descriptions. We consider these complementary strategies aiming to capture different aspects of hadron structure.

The paper is organized as follows. The bridge between EMT and TMDs is discussed in Section \ref{Section2}. The outline of phenomenological models used in the present analysis is presented in Section \ref{Section3}. We present our results and their discussion in Section \ref{Section4}. Finally, we conclude our key results in Section \ref{Section5}.
\section{Transverse momentum-dependent parton distributions and Energy momentum tensor \label{Section2}}
Consider a spin-1/2 target of proton with mass $M$, carrying spin $\lambda=$ ($\Uparrow,\Downarrow$), representing ($\frac{1}{2},-\frac{1}{2}$) and moving with four-momentum $P$ which has light-front coordinates as follows
\begin{eqnarray}
	P&=&\biggl[P^+,\frac{M^2}{2P^+},\textbf{0}_\perp \biggr].
\end{eqnarray}
The four-momentum $k$ of an active quark with mass $m_q$ and carrying longitudinal momentm fraction $x$ is defined as
\begin{eqnarray}
	k&=&\biggl[xP^+,\frac{k^2+\bfk^2}{2xP^+},\bfk \biggr].
\end{eqnarray}
Following the Wilson line based decomposition of gauge field into the  pure and physical components and adopting the light-front gauge $A^+=0$ with appropriate boundary condition, the Wilson line $\mathbb{W}$ reduces to unity, the four-momentum of an active quark can be defined by $k^\alpha=i \partial^\alpha$, i.e., the canonical momentum and the light-front gauge-invariant canonical (gic) EMT operator is given by
\begin{eqnarray}
	T_{q,gic}^{\alpha \beta}(r)=\bar{\Psi}(r)~ \gamma^\alpha \frac{i}{2} \overleftrightarrow{D}^\beta_{pure} \Psi(r),
\end{eqnarray}
where $\overleftrightarrow{D}^\beta_{pure}=\partial^\alpha-igA^\alpha_{pure}$ is termed as pure-gauge covariant derivative and the quark bilocal gic EMT operator is written as
\begin{eqnarray}
	T_{q,gic}^{\alpha \beta}(r,k)  &=& k^\beta \int \frac{d^4z}{(2\pi)^4} e^{i k \cdot z} ~\bar{\Psi}\bigg(r-\frac{z}{2}\bigg)~ \gamma^\alpha~ \mathbb{W} \bigg(r-\frac{z}{2},r+\frac{z}{2}\bigg|n\bigg)~  \Psi\bigg(r+\frac{z}{2} \bigg).
\end{eqnarray}
This well defined form arises because pure-gauge covariant derivatives commute, eliminating gauge-dependent ambiguities in the bilocal operator and the integration of $T^{\alpha \beta}_{q,gic} (r,k)$ over the four-momentum of an active quark gives $T_{q,gic}^{\alpha \beta}(r)$ \cite{Lorce:2012ce, Lorce:2012rr, Chen:2008ag, Wakamatsu:2010cb, Hatta:2011zs, Hatta:2011ku}. The four-momentum $k$ is defined as the Fourier conjugate to the space-time distance $z$ between the quark operators $\bar \Psi$ and $\Psi$. $\gamma^\alpha$ denotes the generic matrix in Dirac space. The full unintegrated EMT $\Theta^{\alpha \beta}_{q }(P,k,N,\lambda;\eta)$ in terms of the forward matrix elements of the operator $T^{\alpha \beta}_{q,gic} (r,k)$ is expressed as
\begin{eqnarray}
	\Theta^{\alpha \beta}_{q } (P,k,N,\lambda;\eta) = \frac{1}{2} \bra{P,\lambda} T^{\alpha \beta}_{q,gic} (0,k) \ket{P,\lambda}. \label{EMT}
\end{eqnarray}
The above mentioned correlator depends on the rescaling-invariant four-vector $N$ and the parameter $\eta$ is defined through the zeroth component of light-front vector $n$, where $n=[0,\eta,\textbf{0}_\perp]$. The integration over the light-front energy of an active quark space leads to the three-dimensional TMD EMT 
\begin{eqnarray}
	\mathbb{T}^{\alpha \beta}_q (P,x,\bfk,N,\lambda;\eta) &=& \int dk^- \Theta^{\alpha \beta}_q (P,k,N,\lambda;\eta) \nonumber \\ &=& \frac{1}{2} \int \frac{dz^- d^2 \bfz}{(2\pi)^3} e^{i k \cdot z} i \partial^\beta_z \bra{P,\lambda}  \bar{\Psi}\bigg(\frac{-z}{2}\bigg) \gamma^\alpha \mathbb{W} \bigg(\frac{-z}{2},\frac{z}{2}\bigg|n\bigg)  \Psi\bigg(\frac{z}{2} \bigg) \ket{P,\lambda} \bigg|_{z^+=0}, \nonumber\\ \label{EMT_TMD}
\end{eqnarray}
that interprets the three-dimensional distribution of the active quark in the momentum space. The fundamental components for parametrizing the TMD EMT are $g^{\alpha \beta}_T$, $\epsilon^{\alpha \beta}_T$, $P^\alpha$, $N^\alpha$ and $k^\alpha_T$. Using the complete set of independent rank-2 tensors, EMT in momentum space \cite{Lorce:2023zzg} is expressed as
\begin{eqnarray}
	\mathbb{T}^{\alpha \beta}_q &=& \frac{1}{P^+} \bigg[ P^\alpha P^\beta a_1 + N^\alpha N^\beta a_2 + \textbf{k}^\alpha_T \textbf{k}^\beta_T a_3 +P^\alpha N^\beta a_4 + N^\alpha P^\beta a_5 + P^\alpha \textbf{k}^\beta_T a_6 + \textbf{k}^\alpha_T P^\beta a_7 +N^\alpha \textbf{k}^\beta_T a_8 \nonumber \\ &+& \textbf{k}^\alpha_T N^\beta a_9 + M^2 g^{\alpha \beta}_T a_0 - \frac{\epsilon^{\textbf{k}_T \textbf{S}_T}}{M} \bigg\{ P^\alpha P^\beta a_{1T}^\perp + N^\alpha N^\beta a_{2T}^\perp + \textbf{k}^\alpha_T \textbf{k}^\beta_T a_{3T}^\perp + P^\alpha N^\beta a_{4T}^\perp + N^\alpha P^\beta a_{5T}^\perp \nonumber \\ &+& P^\alpha \textbf{k}^\beta_T a_{6T}^\perp + \textbf{k}^\alpha_T P^\beta a_{7T}^\perp + N^\alpha \textbf{k}^\beta_T a_{8T}^\perp + \textbf{k}^\alpha_T N^\beta a_{9T}^\perp + M^2 g^{\alpha \beta}_T a_{0T}^\perp \bigg\} - M \big\{ P^\alpha \epsilon^{\beta \textbf{S}_T}_T a_{1T} + P^\beta \epsilon^{\alpha \textbf{S}_T} a_{2T} \nonumber \\ &+& N^\alpha \epsilon^{\beta \textbf{S}_T}_T a_{3T} + N^\beta \epsilon^{\alpha \textbf{S}_T}_T a_{4T} + \textbf{k}^\alpha_T \epsilon^{\beta \textbf{S}_T}_T a_{5T} + \textbf{k}^\beta_T \epsilon^{\alpha \textbf{S}_T}_T a_{6T} \big\} - \lambda \big\{ P^\alpha \epsilon^{\beta \textbf{k}_T}_T a_{1L} + P^\beta \epsilon^{\alpha \textbf{k}_T}_T a_{2L} \nonumber \\ &+& N^\alpha \epsilon^{\beta \textbf{k}_T}_T a_{3L} + N^\beta \epsilon^{\alpha \textbf{k}_T}_T a_{4L} + \textbf{k}^\alpha_T \epsilon^{\beta \textbf{k}_T}_T a_{5L} + \textbf{k}^\beta_T \epsilon^{\alpha \textbf{k}_T}_T a_{6L} \big\} \bigg],
\end{eqnarray}
with $k_T^2=-\bfk^2$ in the vector notation \cite{Lorce:2023zzg}, $\mathbb{T}^{\alpha \beta}_q$ is the compact notation for $\mathbb{T}^{\alpha \beta}_q (P,x,\bfk,N,\lambda;\eta)$ and the standard vector TMD correlator (for $\Gamma=\gamma^\alpha$) is defined as 
\begin{eqnarray}
	\Phi^{[\gamma^\alpha]}_q (P,x,\bfk,N,\lambda;\eta)  &=& \frac{1}{2} \int \frac{dz^- d^2 \bfz}{(2\pi)^3} e^{i k \cdot z}~\bra{P,\lambda} \bar{\Psi}~\bigg(\frac{-z}{2}\bigg) \gamma^\alpha \mathbb{W} \bigg(\frac{-z}{2},\frac{z}{2}\bigg|n\bigg)  \Psi\bigg(\frac{z}{2} \bigg) \ket{P,\lambda} \bigg|_{z^+=0}. \nonumber \\ \label{TMD}
\end{eqnarray}
Hereafter, $\Phi^{[\gamma^\alpha]}_q (P,x,\bfk,N,\lambda;\eta)$ is denoted compactly by $\Phi^{[\gamma^\alpha]}_q$. TMDs from Dirac projections $\gamma^+$, $\gamma^{i}_T$ and $\gamma^-$ are parametrized in terms of respective twist-2, twist-3 and twist-4 as
\begin{eqnarray}
	\Phi^{[\gamma^+]}_q &=& f_1^q - \frac{\epsilon^{\textbf{k}_T \textbf{S}_T}_T}{M} f_{1T}^{\perp(q)}, \label{Twist2} \\
	\Phi^{[\gamma^i_T]}_q &=& \frac{M}{P^+}~ \bigg[\frac{\textbf{k}^i_T}{M} f^{\perp(q)}-\epsilon^{i\textbf{S}_T}_T f_T^{\prime(q)}-\lambda_l \frac{\epsilon^{i\textbf{k}_T}}{M} f_L^{\perp(q)} - \frac{\textbf{k}^i_T \textbf{k}^j_T-\frac{1}{2} \textbf{k}_T^2 g^{ij}_T}{M^2} \epsilon_{Tj\textbf{S}_T} f_T^{\perp(q)} \bigg], \label{Twist3} \\
	\Phi^{[\gamma^-]}_q &=& \bigg(\frac{M}{P^+}\bigg)^2 \bigg[f_3^q - \frac{\epsilon^{\textbf{k}_T \cdot \textbf{S}_T}_T}{M} f_{3T}^{\perp(q)} \bigg], \label{Twist4}
\end{eqnarray}
where the quantity  $\lambda_l$ represents the  longitudinal light-front polarization \cite{Bacchetta:2006tn, Meissner:2009ww} and $i$ index denotes the transverse component. From Eqs. (\ref{EMT_TMD}) and (\ref{TMD}), one can relate the EMT in momentum space with the standard TMD correlator by
\begin{eqnarray}
	\mathbb{T}^{\alpha \beta}_q  &=& k^\beta~ \Phi^{[\gamma^\alpha]}_q,
\end{eqnarray}
for $\beta\neq-$. The quark TMD correlator $\Phi^{[\gamma^+]}_q$ encodes the probability density of finding an active quark with longitudinal momentum fraction and intrinsic transverse momentum, while $\mathbb{T}^{++}_q$ and $\mathbb{T}^{+i}_q$ give the corresponding longitudinal and transverse quark momentum densities from the EMT as
\begin{eqnarray}
	\mathbb{T}^{++}_q &=& xP^+ \Phi^{[\gamma^+]}_q = \biggl(f_1^q - \frac{\epsilon^{\textbf{k}_T \textbf{S}_T}_T}{M} f_{1T}^{\perp(q)} \biggr) xP^+ , \label{T++} \\
	\mathbb{T}^{+i}_q &=& \textbf{k}^i_T \Phi^{[\gamma^+]}_q = \biggl(f_1^q - \frac{\epsilon^{\textbf{k}_T \textbf{S}_T}_T}{M} f_{1T}^{\perp(q)} \biggr) \textbf{k}^i_T. \label{T+i}
\end{eqnarray}
The components of EMT are anti-symmetric ($\mathbb{T}^{+i}_q \neq \mathbb{T}^{i+}_q$), as velocity and canonical momentum are often found not to be parallel. Therefore, the component $\mathbb{T}^{i+}_q$ represents the transverse flux of inertia and can be written in terms of the standard TMDs as
\begin{eqnarray}
	\mathbb{T}^{i+}_q &=& x P^+ \Phi^{[\gamma^i_T]}_q  = \bigg(xf^{\perp(q)} -\frac{\epsilon^{\textbf{k}_T \textbf{S}_T}_T}{M} xf_T^{\perp(q)} \bigg) \textbf{k}^i_T-M \epsilon^{i \textbf{S}_T}_T xf^{+(q)}_T - \lambda_l \epsilon^{i \textbf{k}_T}_T xf^{\perp(q)}_L.
\end{eqnarray}
On decomposition of $\epsilon^{i \textbf{S}_T}_T$ into the parallel and orthogonal components to the transverse components of an active quark momentum, the above equation becomes
\begin{eqnarray}
	\mathbb{T}^{i+}_q = \bigg(xf^{\perp(q)} -M\frac{\epsilon^{\textbf{k}_T \textbf{S}_T}_T}{\textbf{k}^2_T} xf_T^{-(q)} \bigg) \textbf{k}^i_T - \bigg( \lambda_l  xf^{\perp(q)}_L+ \frac{M(\textbf{k}_T\cdot\textbf{S}_T)}{\textbf{k}^2_T} xf^{+(q)}_T \bigg) \epsilon_T^{i \textbf{k}_T}. \label{Ti+}
\end{eqnarray}
 The expression of $f^{\pm(q)}_T$ TMD is
\begin{eqnarray}
	f^{\pm(q)}_T=f^{\prime(q)}_T\pm \frac{\bfk^2}{2M^2} f_T^{\perp(q)}.
\end{eqnarray}
The transverse components of $\mathbb{T}^{\alpha \beta}_q$ gives the transverse pressure $\sigma^q$ and shear force $\Pi^q$ distributions in the momentum space that can be computed by 
\begin{eqnarray}
	\mathbb{T}^{ij}_q &=& -g^{ij}_T \sigma^q +\bigg(\frac{1}{2} g^{ij}_T -\frac{\textbf{k}^i_T \textbf{k}^j_T}{\textbf{k}^2_T} \bigg) \Pi^q + \frac{\textbf{k}^i_T \epsilon^{j \textbf{k}_T}_T + \textbf{k}^j_T \epsilon^{i \textbf{k}_T}_T}{2\textbf{k}^2_T} \Pi_S^q + \epsilon^{ij} _T \Pi_A^q, \label{Tiq}
\end{eqnarray}
which is analogous to the representation of 2D spatial distributions of pressure and shear force \cite{Lorce:2018egm}. On replacing transverse momentum $\bfk$ with conjugate position $\bfb$ by Fourier transformation, first two terms can also be obtained in the position space. The last two transverse tensors are naively T-odd, provided that $\Pi^q_S$ and $\Pi^q_A$ are linear with respect to the target polarization \cite{Lorce:2023zzg}. Considering the TMD EMT, we have
\begin{eqnarray}
	\mathbb{T}^{ij}_q &=& k^j_T ~ [\Phi^{\gamma^i_T}] = \frac{1}{P^+} ~ \bigg[\biggl\{f^{\perp(q)} - \frac{M \epsilon^{\textbf{k}_T \textbf{S}_T}_T}{\textbf{k}^2_T}f^{-(q)}_T\biggr\} - \biggl\{\lambda_l f_L^{\perp(q)} + \frac{M(\textbf{k}_T \cdot \textbf{S}_T)}{\textbf{k}^2_T} f^{+(q)}_T \biggr\}\bigg],
\end{eqnarray}
with 
\begin{eqnarray}
	2 \sigma^q &=&  -\frac{1}{P^+} \bigg[\textbf{k}^2_T ~f^{\perp(q)}-M \epsilon^{\textbf{k}_T \textbf{S}_T}_T f^{-(q)}_T\bigg], \label{sigma}\\
	2 \Pi_A^q &=&  -\frac{1}{P^+} \bigg[\lambda_l~ \textbf{k}^2_T ~f^{\perp(q)}_L+M (\textbf{k}_T \cdot \textbf{S}_T) f^{+(q)}_T\bigg]. \label{Pi}
\end{eqnarray}
Also, $2 \sigma^q = \Pi^q$ and $2 \Pi_A^q=\Pi_S^q$ \cite{Lorce:2023zzg}. Therefore, in momentum-space, pressure and shear force are found to be qualitatively same as they are governed by same twist-3 TMDs. 

For a quark-diquark system of proton, the two-particle Fock state expansion for the case of scalar $\mathfrak{s}$ and axial-vector $\mathfrak{a}$ diquark in terms of LFWFs $\psi^{\Uparrow,\Downarrow}_{\lambda_q}$ and $\psi^{\Uparrow,\Downarrow}_{\lambda_q, \lambda_{\mathfrak{a}}}$ is expressed as
\begin{eqnarray}
	\ket{q\mathfrak{s}}^{\lambda} &=& \int \frac{dx~d^2\bfk}{16 \pi^3 \sqrt{x(1-x)}} \sum_{\lambda_q} \psi^{\lambda}_{\lambda_q}(x,\bfk) \ket{\lambda_q,xP^+,\bfk}, \\
	\ket{q\mathfrak{a}}^{\lambda} &=& \int \frac{dx~d^2\bfk}{16 \pi^3 \sqrt{x(1-x)}} \sum_{\lambda_q, \lambda_{\mathfrak{a}}} \psi^{\lambda}_{\lambda_q, \lambda_{\mathfrak{a}}}(x,\bfk) \ket{\lambda_q,\lambda_{\mathfrak{a}},xP^+,\bfk},
\end{eqnarray}
respectively. Here, $\lambda_q$ ($\lambda_\mathfrak{a}$) denotes the helicity of an active quark (axial-vector diquark). An axial-vector diquark, which is further categorized into isoscalar ($ud$) and isovector ($dd$) diquarks, for respective $u$ and $d$ active quark flavors, is labeled as $\mathfrak{a}_u$ and $\mathfrak{a}_d$. Henceforth, $\mathfrak{a}=$ ($\mathfrak{a}_u, \mathfrak{a}_d$) depending on the flavor of an active quark. Since $\lambda$ (as well as $\lambda_q$) in the initial and final state can have feasible combinations of ($\Uparrow \Uparrow,\Uparrow \Downarrow, \Downarrow \Downarrow, \Downarrow \Uparrow$), we distinguish between the initial and final state helicities of both proton and an active quark by introducing a prime notation to the final state. The quark-quark correlator expansion in terms of the overlap form of LFWFs for scalar and axial-vector diquarks can be written as
\begin{eqnarray}
	\Phi^{[\gamma^\alpha](\lambda,\lambda^\prime)}_{q \mathfrak{s}} &=&\frac{1}{16 \pi^3} \sum_{\lambda_{q}} \sum_{\lambda_{q}^\prime} \psi^{\lambda^\prime}_{\lambda_{q}^\prime}(x,\bfk)~ \psi^{\lambda}_{\lambda_{q}}(x,\bfk)  \frac{\bar{u}(k,\lambda_q^\prime) \gamma^\alpha u(k,\lambda_q) }{2xP^+}, \label{Overlap1} \\
	\Phi^{[\gamma^\alpha](\lambda,\lambda^\prime)}_{q \mathfrak{a}} &=&\frac{1}{16 \pi^3} \sum_{\lambda_{q}} \sum_{\lambda_{\mathfrak{a}}} \sum_{\lambda_{q}^\prime} \psi^{\lambda^\prime}_{\lambda_{q}^\prime \lambda_{\mathfrak{a}}}(x,\bfk)~   \psi^{\lambda}_{\lambda_{q}\lambda_{\mathfrak{a}}}(x,\bfk) \frac{\bar{u}(k,\lambda_q^\prime) \gamma^\alpha u(k,\lambda_q) }{2xP^+}, \label{Overlap2}
\end{eqnarray}
respectively, and their relation with the standard set of leading and higher-twist TMDs has been given in the Table \ref{TableTMDs} with respect to the initial and final state helicities of the proton \cite{Sharma:2024lal}. In order to evaluate the T-odd TMDs, it is necessary to introduce a kernel that encodes the ISI/FSI in the gauge‑link structure of the quark correlator. Physically, this corresponds to accounting for the interactions of the active quark with the target remnant either before the hard scattering (ISI) or after it (FSI). These soft gluon exchanges between the active quark and the residual diquark system are collectively described as gluon rescattering. Therefore, to evaluate T-odd TMDs, a perturbative Abelian gluon exchange approximation of the gauge link is considered \cite{Brodsky:2002cx, Brodsky:2002rv, Gurjar:2022rcl}, and the kernel to include this effect is given by 
\begin{eqnarray}
	i G(x,\bfq)=\frac{\alpha_s C_f}{\pi \bfq^2}.
\end{eqnarray}
Here, $\alpha_s=0.3$ and $C_f=\frac{4}{3}$ are coupling constants and the color factor, respectively. The momentum carried by the gluon is given by $\bfq=\bfk-\bfk^\prime$. Specifically, for T-odd TMDs, overlap form of LFWFs for respective scalar and axial-vector diquarks are modified as
\begin{eqnarray}
	\Phi^{[\gamma^\alpha](\lambda,\lambda^\prime)}_{q \mathfrak{s}} &=& i \int \frac{ d^2 \bfq}{16 \pi^3} \sum_{\lambda_{q}} \sum_{\lambda_{q}^\prime} \psi^{\lambda^\prime}_{\lambda_{q}^\prime}(x,\bfk)G(x,\bfq)  \psi^{\lambda}_{\lambda_{q}}(x,\bfk^\prime)~\frac{\bar{u}(k,\lambda_q^\prime) \gamma^\alpha u(k^\prime,\lambda_q) }{2xP^+}, \label{Overlap3} \\
	\Phi^{[\gamma^\alpha](\lambda,\lambda^\prime)}_{q \mathfrak{a}} &=& i \int \frac{d^2\bfq}{16 \pi^3} \sum_{\lambda_{q}} \sum_{\lambda_{\mathfrak{a}}} \sum_{\lambda_{q}^\prime} \psi^{\lambda^\prime}_{\lambda_{q}^\prime \lambda_{\mathfrak{a}}}(x,\bfk)  G(x,\bfq)\psi^{\lambda}_{\lambda_{q}\lambda_{\mathfrak{a}}}(x,\bfk^\prime) \frac{\bar{u}(k,\lambda_q^\prime) \gamma^\alpha u(k,\lambda_q) }{2xP^+}. \nonumber\\ \label{Overlap4} 
\end{eqnarray}
\begin{table*}[t]
\centering
\begin{tabular}{|l|c|c|c|c|} 
	\hline
	$\gamma^\alpha$ & $~~\Phi_{q \mathfrak{s}(\mathfrak{a})}^{[\gamma^\alpha](\Uparrow \Uparrow)}+\Phi_{q \mathfrak{s}(\mathfrak{a})}^{[\gamma^\alpha](\Downarrow \Downarrow)}~~$ & $~~\Phi_{q \mathfrak{s}(\mathfrak{a})}^{[\gamma^\alpha](\Uparrow \Uparrow)}-\Phi_{q \mathfrak{s}(\mathfrak{a})}^{[\gamma^\alpha](\Downarrow \Downarrow)}~~$ & $~~\Phi_{q \mathfrak{s}(\mathfrak{a})}^{[\gamma^\alpha](\Downarrow \Uparrow)}+\Phi_{q \mathfrak{s}(\mathfrak{a})}^{[\gamma^\alpha](\Uparrow \Downarrow)}~~$ & $~~\Phi_{q \mathfrak{s}(\mathfrak{a})}^{[\gamma^\alpha](\Downarrow \Uparrow)}-\Phi_{q \mathfrak{s}(\mathfrak{a})}^{[\gamma^\alpha](\Uparrow \Downarrow)}~~$ \\
	\hline
	$\gamma^+$ & 2$f_1^q$ & - & $\frac{-2 i k_2}{M} f_{1T}^{\perp(q)}$ & $\frac{2k_1}{M}f_{1T}^{\perp(q)}$ \\ \hline 
	$\gamma^1_T$ & $\frac{2k_1}{P^+}f^{\perp(q)}$ & $\frac{-2 i k_2}{P^+} f_{L}^{\perp(q)}$  & $\frac{2 i k_1 k_2}{M P^+} f_T^{\perp(q)}$ & $\frac{-2 M}{P^+} \bigg(f_T^{\prime(q)} + \frac{k^2_1}{M^2} f_T^{\perp(q)}\bigg)$ \\ \hline
	$\gamma^2_T$ & $\frac{2k_2}{P^+}f^{\perp(q)}$ & $\frac{2 i k_1}{P^+} f_{L}^{\perp(q)}$ & $\frac{2i M}{P^+} \bigg(f_T^{\prime(q)} + \frac{k^2_2}{M^2} f_T^{\perp(q)}\bigg)$ & $\frac{-2 k_1 k_2}{M P^+} f_T^{\perp(q)}$ \\ \hline
	$\gamma^-$ & $\frac{2M^2}{(P^+)^2} f_3^q$ & - & $\frac{-2 i k_2 M}{(P^+)^2} f_{3T}^{\perp(q)}$ & $\frac{2k_1 M}{(P^+)^2}f_{3T}^{\perp(q)}$ \\ 
	\hline
\end{tabular}
	\caption{Parametrization of the leading and higher-twist TMDs $\Phi^{\gamma^\alpha}_q$ as mentioned in Eqs. (\ref{Twist2}-\ref{Twist4}), for different combinations of $\lambda$ in initial and final state ($\Uparrow \Uparrow,\Uparrow \Downarrow, \Downarrow \Downarrow, \Downarrow \Uparrow$) \cite{Sharma:2024lal}.}
\label{TableTMDs}
\end{table*}
\section{Model description \label{Section3}}
Given the general TMD operator definitions, the physical content of the EMT is determined by the choice of LFWFs that enter into the TMD quark-quark correlator Eqs. (\ref{Overlap1}-\ref{Overlap4}).
Both models are based on the consideration that every interaction the proton is going to involve, is occurring through the active quark, whereas the remaining part of proton acts as a spectator, therefore it includes the probability of running into every possible active quark-spectator combination. The proton's spin-flavor structure is designed by an amalgam of isoscalar-scalar diquark singlet $|u~ \mathfrak{s}(ud)\rangle$, isoscalar-vector diquark $|u~ \mathfrak{a}_u(ud)\rangle$, and isovector-vector diquark $|d~ \mathfrak{a}_d (uu)\rangle$ states as
\begin{equation}
	|P; \pm \rangle = \sum C_\mathfrak{d} |q~ \mathfrak{d} \rangle^\pm, \label{PS_state}
\end{equation}
where $\mathfrak{d}=(\mathfrak{s}, \mathfrak{a}_u, \mathfrak{a}_d)$ denotes the different types of diquarks.
%
%


\subsection{Model-1 (vertex‑based, Spectator diquark model)}
In spectator diquark model, the LFWFs are computed by considering the dipolar form factor at the proton-quark-diquark vertex over the pointlike form
to prevent divergence. Different choices are available in the literature while summing over the complete set of axial-vector diquark polarization states. But, we have considered only the physical lonitudinal and transversely polarized axial-vector diquarks, along with scalar diquark, dropping out the time-like unphysical polarization state of an axial-vector diquark. The LFWFs for the respective scalar diquark $\mathfrak{s}$ with mass $m_\mathfrak{s}$ and axial-vector diquarks $\mathfrak{a}$ with mass $m_\mathfrak{a}$ are expressed as
\begin{eqnarray}
	\psi^{\lambda}_{\lambda_q} (x,\bfk) &=& \sqrt{\frac{k^+}{(P-k)^+}} \frac{1}{k^2-m^2_q} \bar{u} (k,\lambda_q) ~\mathcal{Y}_{\mathfrak{s}}~ U(P,\lambda), \\
	\psi^{\lambda}_{\lambda_q \lambda_{\mathfrak{a}}}  (x,\bfk) &=& \sqrt{\frac{k^+}{(P-k)^+}} \frac{1}{k^2-m^2_q} \bar{u} (k,\lambda_q) \epsilon^\ast_\mu (P-k,\lambda_{\mathfrak{a}}) \cdot \mathcal{Y}_{\mathfrak{a}}^\mu~ U(P,\lambda),
	\label{VectorWfn}
	\label{ScalarWfn}
\end{eqnarray}
with  scalar and axial-vector vertex $\mathcal{Y}_{\mathfrak{s}}=ig_{\mathfrak{s}}(k^2) \bf{1}$ and $\mathcal{Y}_{\mathfrak{a}}^\mu=ig_{\mathfrak{a}}(k^2) \gamma^\mu \gamma_5 /\sqrt{2}$, sequentially. The dipolar form factor for spectator diquarks has a form 
\begin{eqnarray}
	g_{\mathfrak{d}}(k^2) = g_{\mathfrak{d}} \frac{k^2-m_q^2}{|k^2-\Lambda^2_{\mathfrak{d}}|}.
\end{eqnarray}
Here, the $\Lambda_{\mathfrak{d}}$ and $g_{\mathfrak{d}}$ is the cutoff parameter and coupling constant, respectively. The momentum space wave function in the spectator diquark model is defined as
\begin{eqnarray}
	\phi_{q\mathfrak{d}} =- \frac{g_{\mathfrak{d}}}{\sqrt{1-x}} \frac{x (1-x)^2}{(\bfk^2+L^2_{q\mathfrak{d}})^2} \, ,
	\label{MSWF}
\end{eqnarray}
with $L^2_{q\mathfrak{d}}=x m^2_{\mathfrak{d}}+(1-x) \Lambda_{q\mathfrak{d}}^2-x(1-x)M^2$. Here, 
$\Lambda_{q\mathfrak{d}}$ (GeV) is an appropriate cut-off to avoid the singularity and satisfy the relation $m_{\mathfrak{d}}>M-\Lambda_{q\mathfrak{d}}$. The free parameters of the spectator diquark model are computed using $MINUIT$ program by fitting the normalized $f_1^q$ TMD as detailed in Ref. \cite{Bacchetta:2008af}. 
The original couplings $g_{\mathfrak{d}}$
are rescaled by normalization constants $N_{\mathfrak{d}}$ and the $f_1^q$ TMD is normalized by
\begin{eqnarray}
	\int dx \int d^2 \bfk ~ f^q_{1norm}(x,\bfk) = 1.
\end{eqnarray}
For the sake of completeness, all the free parameters used in the present calculations are mentioned in the Table \ref{TableParameters}. 

\begin{table*}[t]
	\centering
	\begin{tabular}{|l|c|c|c|} 
		\hline
		$\text{Diquark}$ ($\mathfrak{d}$) & $m_{\mathfrak{d}}$ (GeV) & $\Lambda_{\mathfrak{d}}$ (GeV) & $C_{\mathfrak{d}}$ \\
		\hline
		
		$\text{Scalar}~(\mathfrak{s})$ & $0.822\pm0.053$ & $0.609\pm0.038$ & $0.847\pm0.111$ \\ 
		\hline
		$\text{Axial-vector}~(\mathfrak{a}_u)$ & $1.492\pm0.173$  & $0.716\pm0.074$ & $1.061\pm0.085$ \\ 
		\hline
		$\text{Axial-vector}~(\mathfrak{a}_d)$ & $0.890\pm0.008$ & $0.376\pm0.005$ & $0.880\pm0.008$ \\ 
		\hline
	\end{tabular}
	\caption{Free parameters of Specattor diquark model, used in the present calculation for quark mass $m_q=0.3$ GeV and proton mass $M=0.9$ GeV.}
	\label{TableParameters}
\end{table*}
\subsection{Model-2 (AdS/QCD inspired, Light-front quark-diquark model)}
The standard form of the LFWFs $\varphi^{(\nu)}_{i}(x,\bfp)$, appeared in Table \ref{LFWFs} is motivated from the soft-wall AdS/QCD \cite{deTeramond:2011aml} predictions as:
	\begin{eqnarray}
		\varphi_\mathfrak{d}^{(q)}(x,\bfp)=\frac{4\pi}{\kappa}\sqrt{\frac{\log(1/x)}{1-x}}x^{a_\mathfrak{d}^q}(1-x)^{b_\mathfrak{d}^q}\exp\Bigg[-\delta^q\frac{\bfp^2}{2\kappa^2}\frac{\log(1/x)}{(1-x)^2}\bigg].
		\label{LFWF_phi}
	\end{eqnarray}
The parameters have been fitted to the model scale $\mu_0=0.313{\ \rm GeV}$ are shown in Table \ref{tab_par_LFQDM} \cite{Maji:2016yqo}. The constituent quark mass \( m_q \) and proton mass $M$ is taken to be $0.055~\mathrm{GeV}$ and $0.938~\mathrm{GeV}$ sequentially \cite{Chakrabarti:2019wjx}.
%
%
\begin{table}[h]
	\centering
	\begin{tabular}{|c|c|c|}
		\hline
		\rule{0pt}{4.7mm}\backslashbox{Parameter}{$\nu$~~}       & $u$                 & $d$                         \\[0.44 mm] \hline
		\rule{0pt}{4.7mm}$C_{\mathfrak{s}}^{2}$ & $1.3872$            & $0$                     \\ \hline
		\rule{0pt}{4.7mm}$C_{\mathfrak{a}_u}^{2}$ & $0.6128$            & $0$                    \\[0.44 mm] \hline
		\rule{0pt}{4.7mm}$C_{\mathfrak{a}_d}^{2}$ & $0$            & $1$                    \\[0.44 mm] \hline
		\rule{0pt}{4.7mm}$N_{S}$     & $2.0191$            & $0$                         \\[0.44 mm] \hline
		\rule{0pt}{4.7mm}$N_0^{\nu}$ & $3.2050$            & $5.9423$                    \\[0.44 mm] \hline
		\rule{0pt}{4.7mm}$N_1^{\nu}$ & $0.9895$            & $1.1616$                    \\[0.44 mm] \hline
		\rule{0pt}{4.7mm}$a_1^{\nu}$ & $0.280\pm 0.001$    & $0.5850 \pm 0.0003$         \\[0.44 mm] \hline
		\rule{0pt}{4.7mm}$b_1^{\nu}$ & $0.1716 \pm 0.0051$ & $0.7000 \pm 0.0002$         \\[0.44 mm] \hline
		\rule{0pt}{4.7mm}$a_2^{\nu}$ & $0.84 \pm 0.02$     & $0.9434^{+0.0017}_{-0.0013}$\\[0.44 mm] \hline
		\rule{0pt}{4.7mm}$b_2^{\nu}$ & $0.2284 \pm 0.0035$ & $0.64^{+0.0082}_{-0.0022}$  \\[0.44 mm] \hline 
		$\delta^\nu$ & $1$ & $1$   \\[0.44 mm] \hline 
        		$\kappa$ & $0.4$ & $0.4$   \\[0.44 mm] \hline 
	\end{tabular}
	\caption{Light-front quark diquark model parameters and constants for both flavors of active quark.}
	\label{tab_par_LFQDM} 
\end{table}
%
The explicit form of the LFWFs corresponding to both models, which are used to compute higher‑twist distributions and to explore the associated mechanical properties in momentum space, are given in Table [\ref{LFWFs}]. 
\begin{table}[h]
	\centering
	\begin{tabular}{|c|c|c|c|c|}
		\hline
		$~~~~~J_z~~~~~$ & $~~~~~\lambda_q~~~~~$ & $~~~~~\lambda_{Sp}~~~~~$ & SDM  & LFQDM \\
		\hline
		\multirow{2}{*}{$~~+\frac{1}{2}~~$} & $~~+\frac{1}{2}~~$ & $-$ & $\frac{m_q+x M}{x} \, \phi_{{q\mathfrak{s}}}$ & $N_\mathfrak{s}~ \varphi^{(q)}_{1}$ \\
		 & $~~-\frac{1}{2}~~$ & $-$ & $-\frac{k_1+i k_2}{x}  \phi_{{q\mathfrak{s}}}$ & $~-N_\mathfrak{s} \frac{k^1+ik^2}{xM}~ \varphi^{(q)}_{2}$ \\
		\hline
		\multirow{6}{*}{$~~+\frac{1}{2}~~$} & $~~+\frac{1}{2}~~$ & $+1$ & $\frac{k_1-i k_2}{x(1-x)} \, \phi_{{q\mathfrak{a}}}$ & $N^{(q)}_1 \sqrt{\frac{2}{3}} \frac{k^1-ik^2}{xM}~  \varphi^{(q)}_{2}$ \\
		 & $~~-\frac{1}{2}~~$ & $+1$ & $\frac{m_q+x M}{x} \, \phi_{{q\mathfrak{a}}}$ & $N^{(q)}_1 \sqrt{\frac{2}{3}}~ \varphi^{(q)}_{1}$ \\
		 & $~~+\frac{1}{2}~~$ & $0$ & $~~~~\frac{\bfk^2-x m_{\mathfrak{a}}^2-m_q M (1-x)^2}{\sqrt{2}x(1-x)m_{\mathfrak{a}}} \, \phi_{{q\mathfrak{a}}}~~~~$ & $-N^{(q)}_0 \sqrt{\frac{1}{3}}~ \varphi^{(q)}_{1}$ \\
		 & $~~-\frac{1}{2}~~$ & $0$ & $\frac{(m_q+M)(k_1+i k_2)}{\sqrt{2}x m_{\mathfrak{a}}}  \phi_{{q\mathfrak{a}}}$ & $~~~~N^{(q)}_0 \sqrt{\frac{1}{3}} \bigg(\frac{k^1+ik^2}{xM} \bigg)~ \varphi^{(q)}_{2}~~~~$ \\
		 & $~~+\frac{1}{2}~~$ & $-1$ & $-x \frac{k_1+i k_2}{x(1-x)}  \phi_{{q\mathfrak{a}}}$ & 0 \\
		 & $~~-\frac{1}{2}~~$ & $-1$ & 0 & 0 \\
		\hline
		\multirow{2}{*}{$~~-\frac{1}{2}~~$} & $~~+\frac{1}{2}~~$ & $-$ & $-\frac{k_1-i k_2}{x}  \phi_{{q\mathfrak{s}}}$ & $~N_\mathfrak{s} \frac{k^1-ik^2}{xM}~ \varphi^{(q)}_{2}$ \\
		& $~~-\frac{1}{2}~~$ & $-$ & $\frac{m_q+x M}{x} \, \phi_{{q\mathfrak{s}}}$  & $N_\mathfrak{s}~ \varphi^{(q)}_{1}$ \\
		\hline
		\multirow{6}{*}{$~~-\frac{1}{2}~~$} & $~~+\frac{1}{2}~~$ & $+1$ & 0 & 0 \\
		& $~~-\frac{1}{2}~~$ & $+1$ & $-x \frac{k_1-i k_2}{x(1-x)}  \phi_{{q\mathfrak{a}}}$ & 0 \\
		& $~~+\frac{1}{2}~~$ & $0$ & $\frac{(m_q+M)(k_1-i k_2)}{\sqrt{2}x m_{\mathfrak{a}}}  \phi_{{q\mathfrak{a}}}$ & $~N^{(q)}_0 \sqrt{\frac{1}{3}} \bigg( \frac{k^1-ik^2}{xM} \bigg)~  \varphi^{(q)}_{2}$ \\
		& $~~-\frac{1}{2}~~$ & $0$ & $-\frac{\bfk^2-x m_{\mathfrak{a}}^2-m_q M (1-x)^2}{\sqrt{2}x(1-x)m_{\mathfrak{a}}} \, \phi_{{q\mathfrak{a}}}$ & $~N^{(q)}_0\sqrt{\frac{1}{3}}~  \varphi^{(q)}_{1}$ \\
		& $~~+\frac{1}{2}~~$ & $-1$ & $-\frac{m_q+x M}{x} \, \phi_{{q\mathfrak{a}}}$ & $~- N^{(q)}_1 \sqrt{\frac{2}{3}}~  \varphi^{(q)}_{1}$ \\
		& $~~-\frac{1}{2}~~$ & $-1$ & $\frac{k_1+i k_2}{x(1-x)} \, \phi_{{q\mathfrak{a}}}$ & $~N^{(q)}_1 \sqrt{\frac{2}{3}} \bigg(\frac{k^1+ik^2}{xM}\bigg)~  \varphi^{(q)}_{2}$ \\
		\hline
	\end{tabular}
	\caption{The light-front wave functions for a proton carrying helicity $J_z$, comprised of an active quark and a spectator diquark with helicities $\lambda_q$ and $\lambda_{\mathfrak{d}}$, respectively, for two different models, specatator diquark model (SDM) and light-front quark diquark model (LFQDM), used in the present calculations with $\phi_{{q\mathfrak{d}}}$ and $\varphi^{(q)}_{1(2)}$ as momentum space wave functions. $M$, $m_q$ and $m_\mathfrak{a}$ are the masses of proton, active quark and a spectator diquark. $N_\mathfrak{s}$, $N^{(q)}_0$ and $N^{(q)}_1$ are the normalization constants.}
    \label{LFWFs}
\end{table}

\begin{figure*}
		\centering
		\begin{minipage}[c]{0.98\textwidth}
			(a)\includegraphics[width=5.5cm]{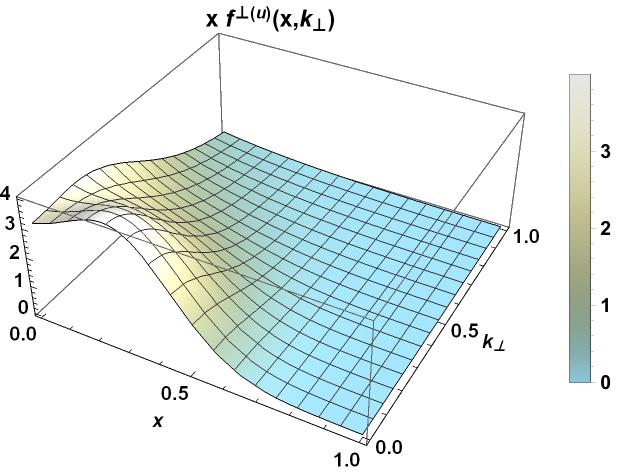}
			\hspace{0.5cm}
			(b)\includegraphics[width=5.5cm]{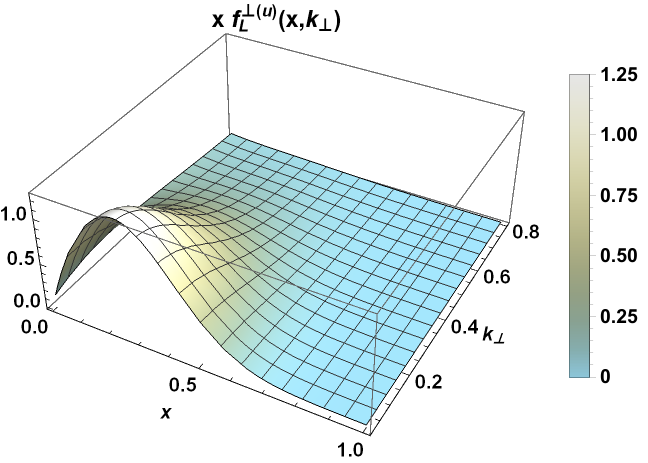}
			\hspace{0.5cm} \\
			(c)\includegraphics[width=5.5cm]{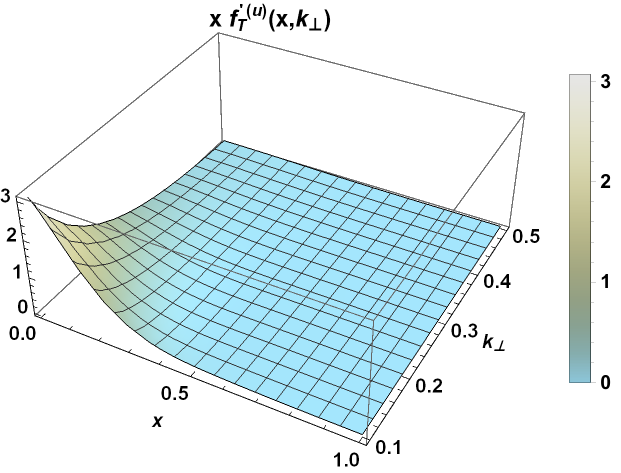}
			\hspace{0.5cm}
			(d)\includegraphics[width=5.5cm]{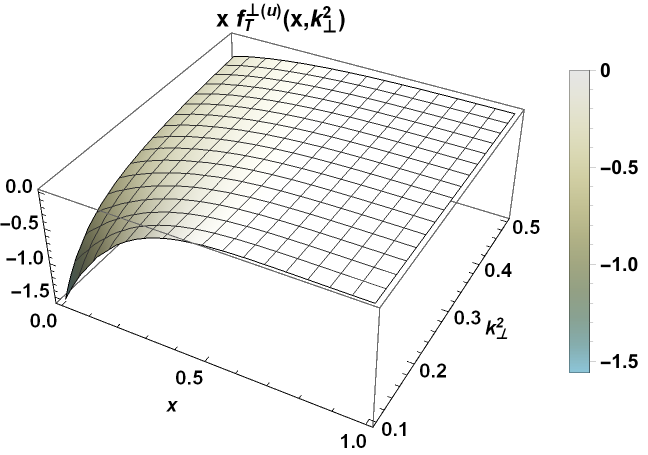}
			\hspace{0.5cm} 
		\end{minipage}
		\caption{\label{fig3du_nk} 
        $x$-weighted T-even and T-odd subleading-twist TMDs: (a) $f^{\perp (u)}(x,k_\perp^2)$, (b) $f_L^{\perp (u)}(x,k_\perp^2)$, (c) $f_T^{\prime (u)}(x,k_\perp^2)$, and (d) $f_T^{\perp (u)}(x,k_\perp^2)$ as functions of the longitudinal momentum fraction $x$ and transverse momentum $\bfk$ (GeV) for the $u$-quark flavor of the proton in SDM.
        }
	\end{figure*} 
\begin{figure*}
		\centering
		\begin{minipage}[c]{0.98\textwidth}
			(a)\includegraphics[width=5.5cm]{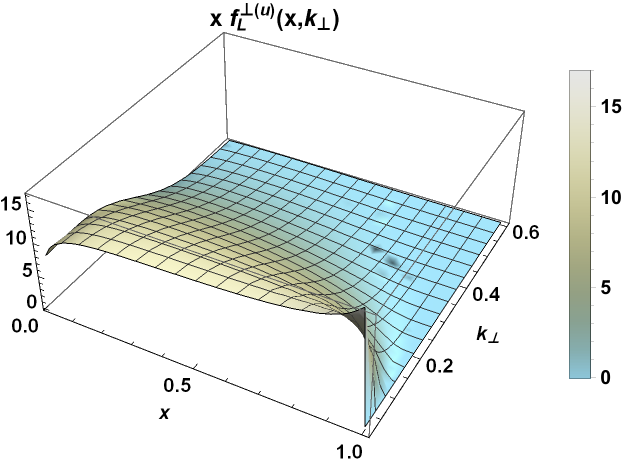}
			\hspace{0.5cm}
			(b)\includegraphics[width=5.5cm]{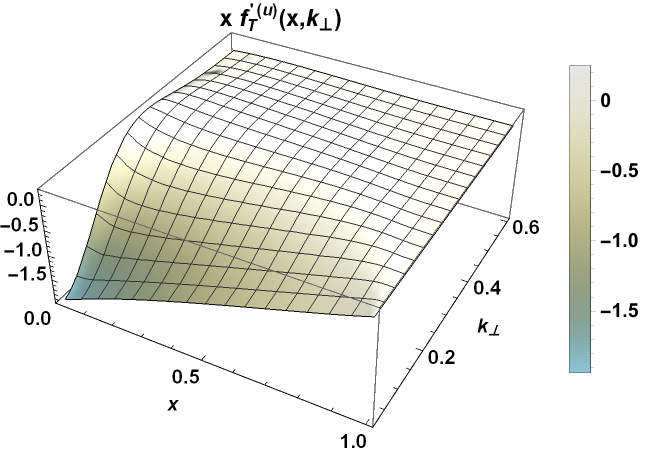}
			\hspace{0.5cm}
		\end{minipage}
		\caption{\label{fig3du_ss} 
        $x$-weighted T-odd subleading-twist TMDs: (a) $f_L^{\perp (u)}(x,k_\perp^2)$ and (b) $f_T^{\prime (u)}(x,k_\perp^2)$ as functions of the longitudinal momentum fraction $x$ and transverse momentum $\bfk$ (GeV) for the $u$-quark flavor of the proton in LFQDM.
		}
	\end{figure*} 
	\begin{figure*}
		\centering
		\begin{minipage}[c]{0.98\textwidth}
			(a)\includegraphics[width=5.5cm]{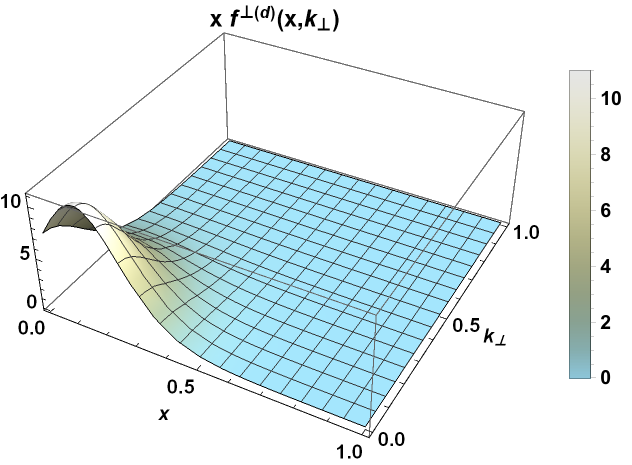}
			\hspace{0.5cm}
			(b)\includegraphics[width=5.5cm]{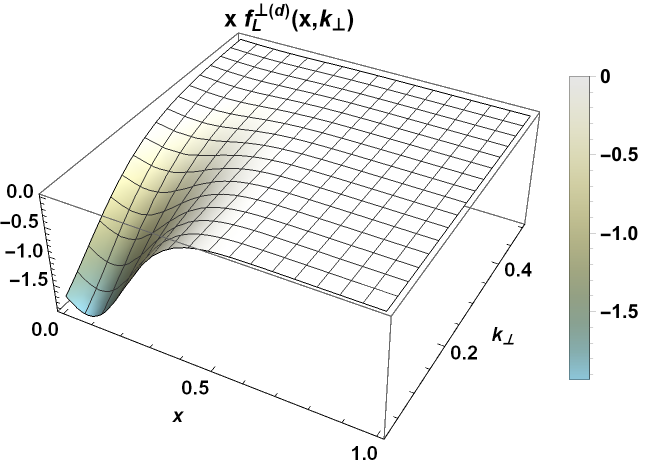}
			\hspace{0.5cm}
			(c)\includegraphics[width=5.5cm]{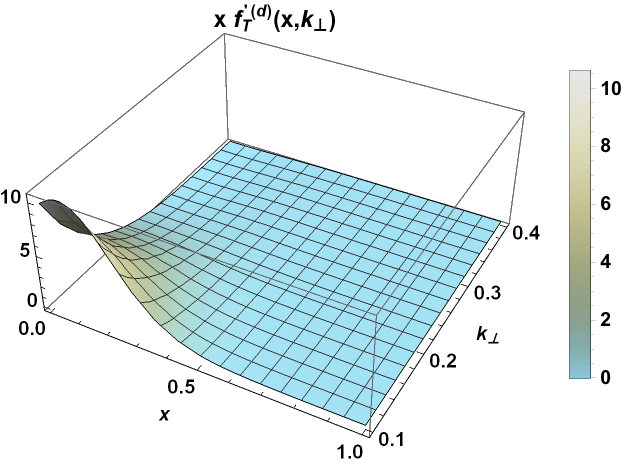}
			\hspace{0.5cm}
			(d)\includegraphics[width=5.5cm]{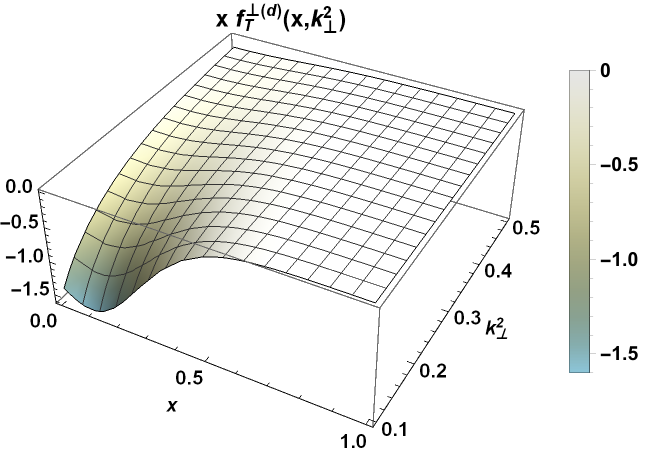}	 
			\hspace{0.5cm}	
		\end{minipage}
		\caption{\label{fig3dd_nk} 
        $x$-weighted T-even and T-odd subleading-twist TMDs: (a) $f_L^{\perp (d)}(x,k_\perp^2)$ and (b) $f_T^{\prime (d)}(x,k_\perp^2)$ as functions of the longitudinal momentum fraction $x$ and transverse momentum $\bfk$ (GeV) for the $d$-quark flavor of the proton in SDM.
		}
	\end{figure*} 
	\begin{figure*}
		\centering
		\begin{minipage}[c]{0.98\textwidth}
			(a)\includegraphics[width=5.5cm]{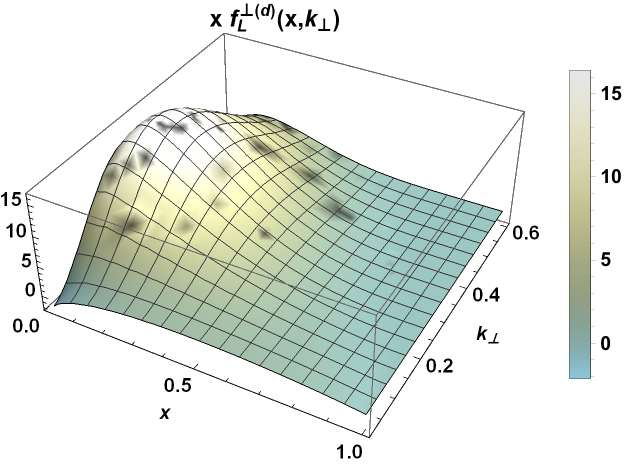}
			\hspace{0.5cm}
			(b)\includegraphics[width=5.5cm]{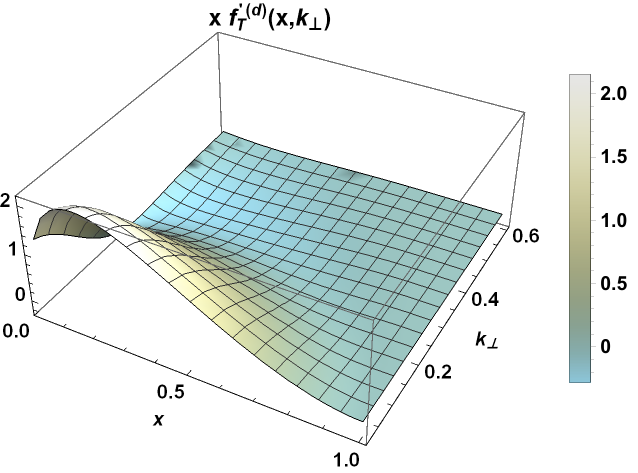}
			\hspace{0.5cm}
		\end{minipage}
		\caption{\label{fig3dd_ss} 
        $x$-weighted T-odd subleading-twist TMDs: (a) $f_L^{\perp (d)}(x,k_\perp^2)$ and (b) $f_T^{\prime (d)}(x,k_\perp^2)$ as functions of the longitudinal momentum fraction $x$ and transverse momentum $\bfk$ (GeV) for the $d$-quark flavor of the proton in LFQDM.
		}
	\end{figure*} 
\begin{figure*}
	\centering
	\begin{minipage}[c]{0.98\textwidth}
		(a)\includegraphics[width=5.5cm]{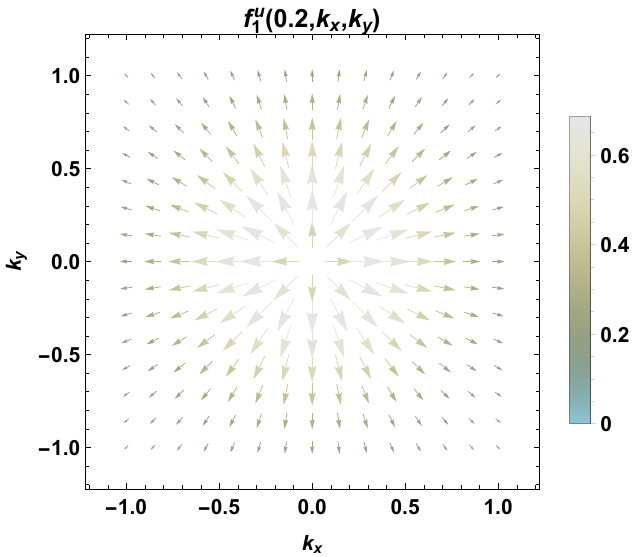}
		\hspace{0.03cm}
		(b)\includegraphics[width=5.5cm]{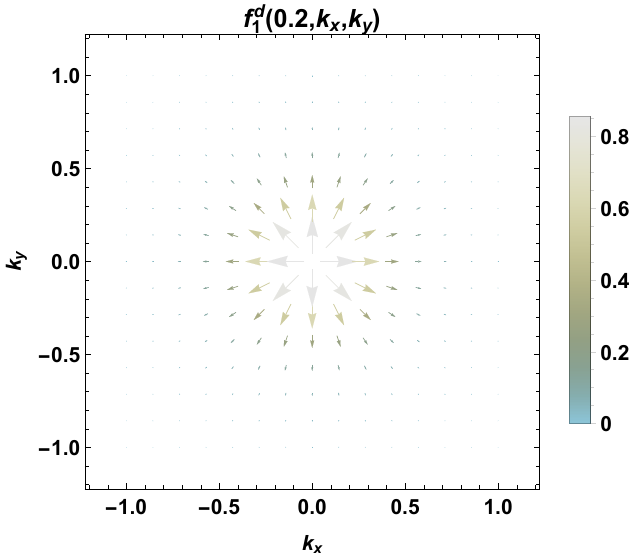}
		\hspace{0.03cm} \\
		(c)\includegraphics[width=5.5cm]{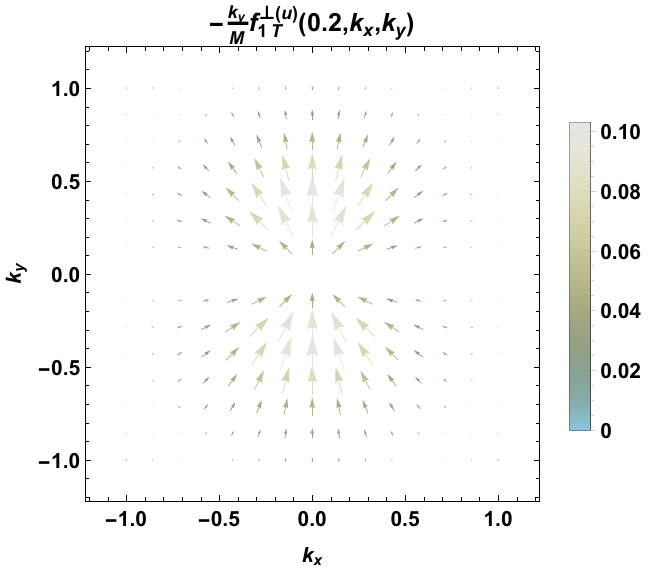}
		\hspace{0.03cm}
		(d)\includegraphics[width=5.5cm]{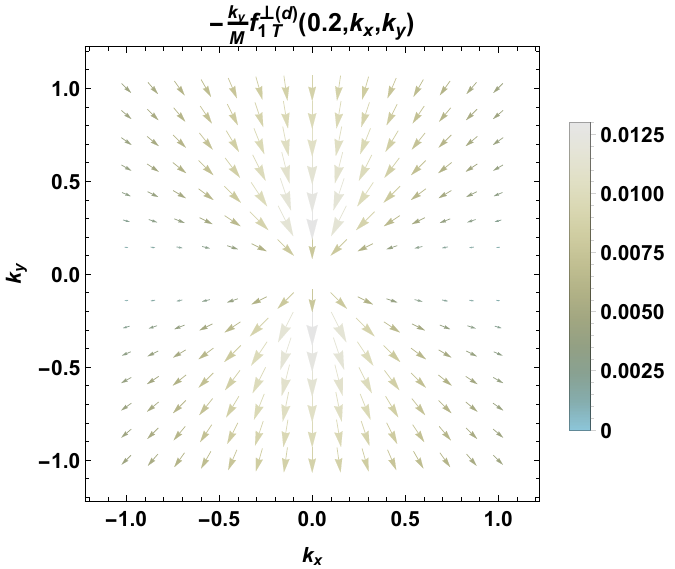}
		\hspace{0.03cm}
	\end{minipage}
	\caption{\label{figtpp_nk} 
    Components of the TMD EMT $T^{++}$: $f_1^q$ and $-\frac{\epsilon_T^{\mathbf{k}_T \mathbf{S}_T}}{M}f_{1T}^{\perp (q)}$ for the $u$- and $d$-quark flavors of the proton in the transverse momentum $(k_x-k_y)$ plane in SDM.
	}
\end{figure*} 
\begin{figure*}
	\centering
	\begin{minipage}[c]{0.98\textwidth}
		(a)\includegraphics[width=5.5cm]{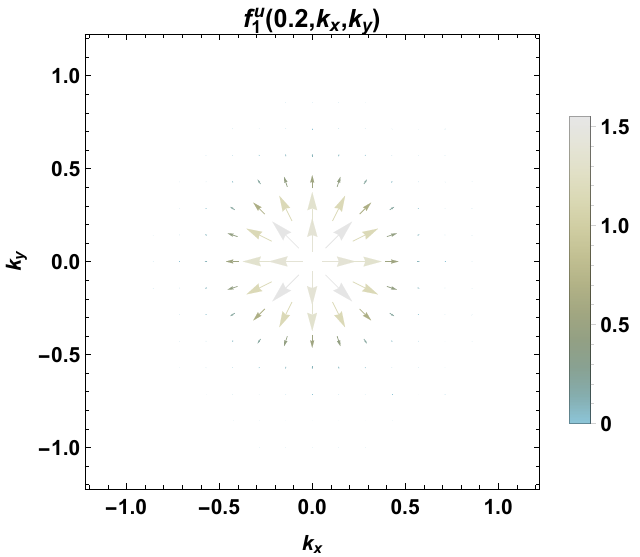}
		\hspace{0.03cm}
		(b)\includegraphics[width=5.5cm]{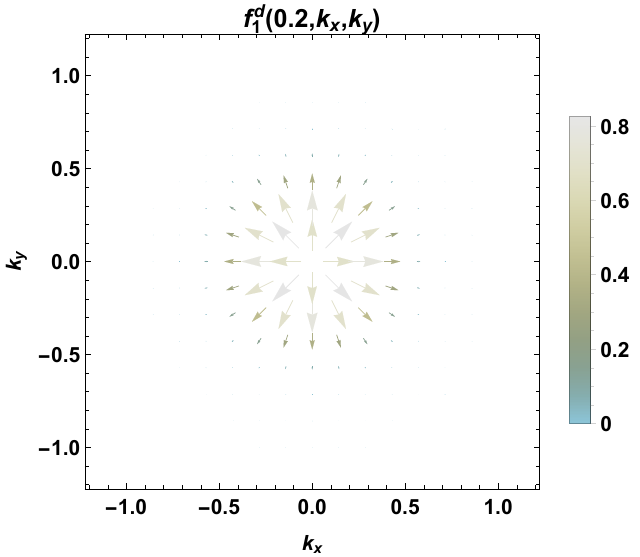}
		\hspace{0.03cm} \\
        (c)\includegraphics[width=5.5cm]{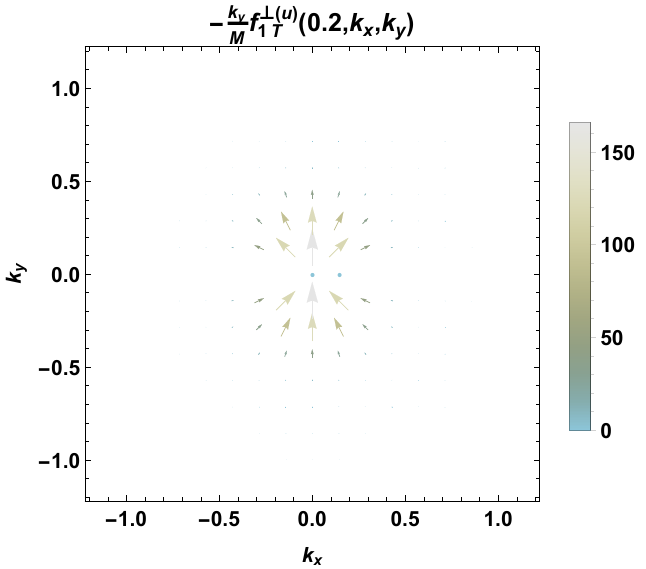}
		\hspace{0.03cm}
		(d)\includegraphics[width=5.5cm]{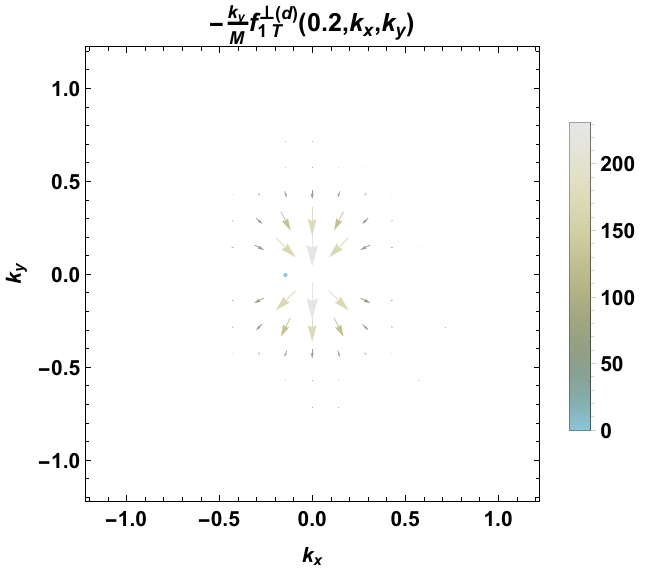}
		\hspace{0.03cm}
	\end{minipage}
	\caption{\label{figtpp_ss} 
    Components of the TMD EMT $T^{++}$: $f_1^q$ and $-\frac{\epsilon_T^{\mathbf{k}_T \mathbf{S}_T}}{M}f_{1T}^{\perp (q)}$ for the $u$- and $d$-quark flavors of the proton in the transverse momentum $(k_x-k_y)$ plane in LFQDM.
	}
\end{figure*} 
\begin{figure*}
	\centering
	\begin{minipage}[c]{0.98\textwidth}
		(a)\includegraphics[width=5.5cm]{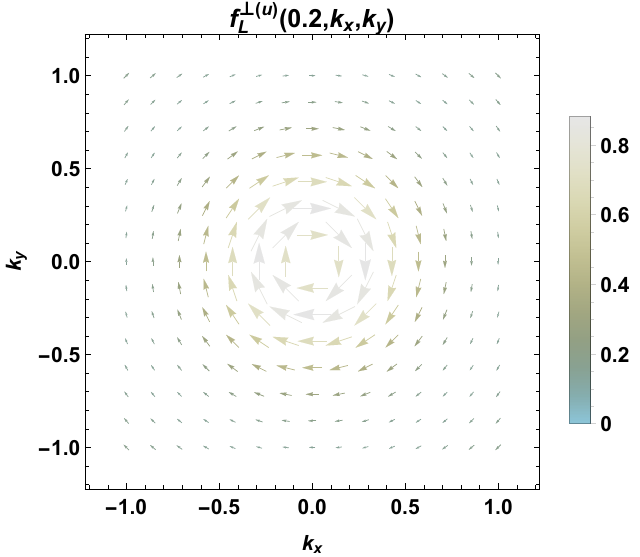}
		\hspace{0.03cm}
		(b)\includegraphics[width=5.5cm]{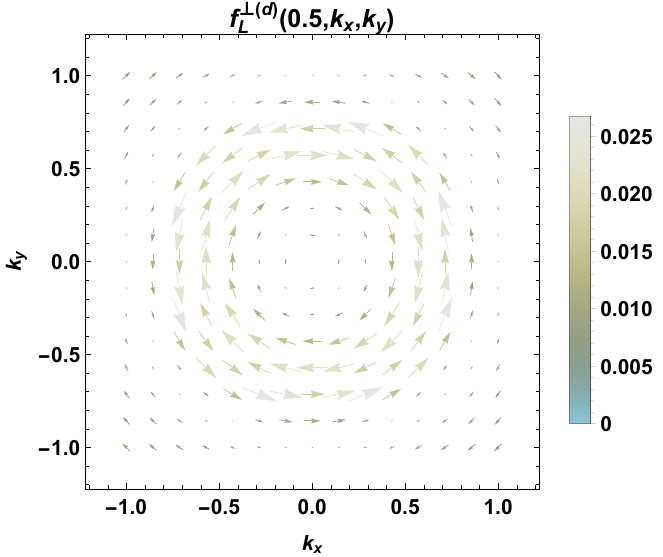}
		\hspace{0.03cm}
	\end{minipage}
	\caption{\label{figtip1_nk} 
    Third component of the TMD EMT $T^{i+}$, $f_{L}^{\perp (q)}$, for (a) $u$- and (b) $d$-quark flavors of the proton in the transverse momentum $(k_x-k_y)$ plane in SDM.
	}
\end{figure*} 
\begin{figure*}
	\centering
	\begin{minipage}[c]{0.98\textwidth}
		(a)\includegraphics[width=5.5cm]{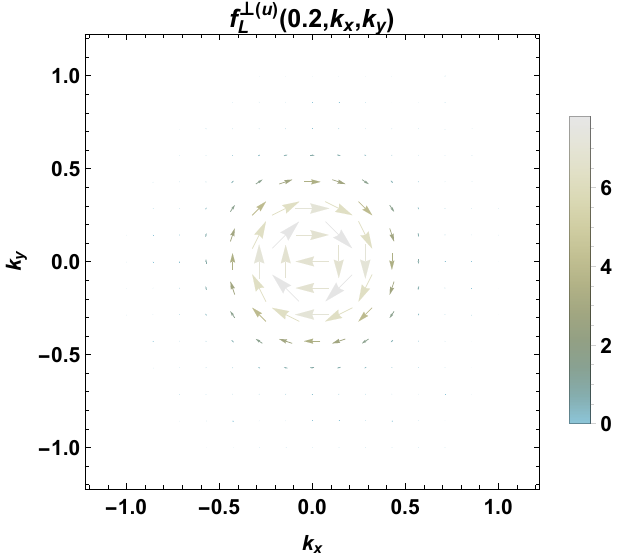}
		\hspace{0.03cm}
		(b)\includegraphics[width=5.5cm]{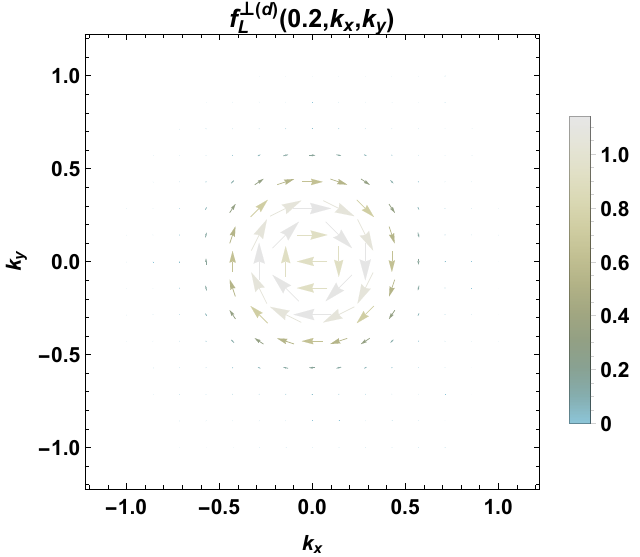}
		\hspace{0.03cm}
	\end{minipage}
	\caption{\label{figtip1_ss} 
    Third component of the TMD EMT $T^{i+}$, $f_{L}^{\perp (q)}$, for (a) $u$- and (b) $d$-quark flavors of the proton in the transverse momentum $(k_x-k_y)$ plane in LFQDM.
	}
\end{figure*} 
\begin{figure*}
	\centering
	\begin{minipage}[c]{0.98\textwidth}
		(a)\includegraphics[width=5.5cm]{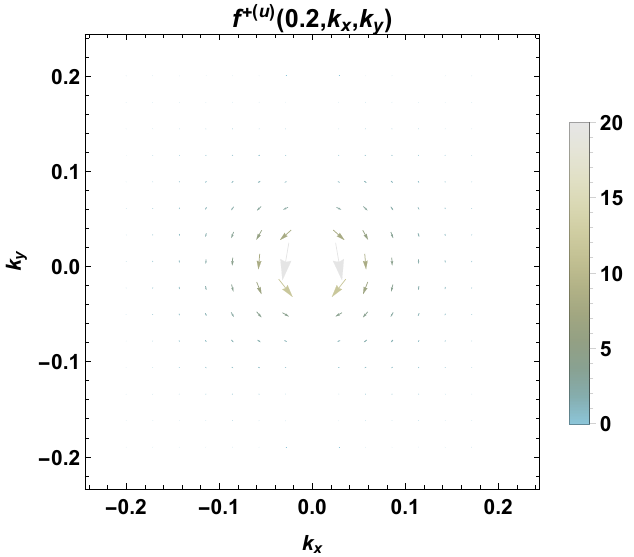}
		\hspace{0.03cm}
		(b)\includegraphics[width=5.5cm]{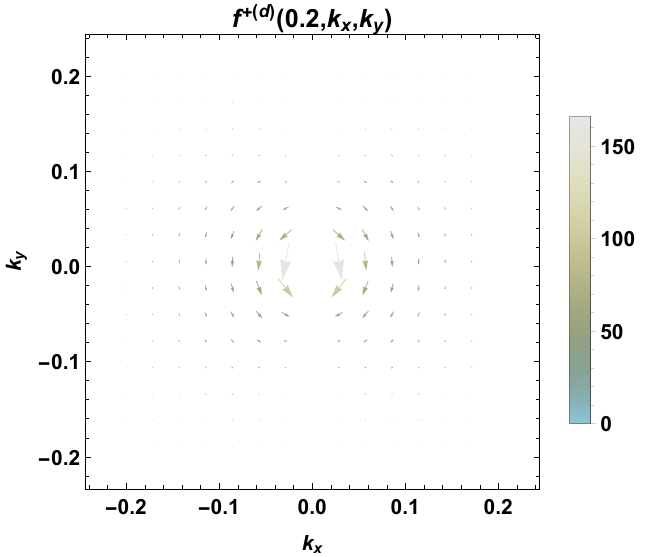}
		\hspace{0.03cm}
	\end{minipage}
	\caption{\label{figtip2_nk} 
    Third component of the TMD EMT $T^{i+}$, $f^{+ (q)}$, for (a) $u$- and (b) $d$-quark flavors of the proton in the transverse momentum $(k_x-k_y)$ plane in SDM.
	}
\end{figure*} 
\begin{figure*}
	\centering
	\begin{minipage}[c]{0.98\textwidth}
		(a)\includegraphics[width=5.5cm]{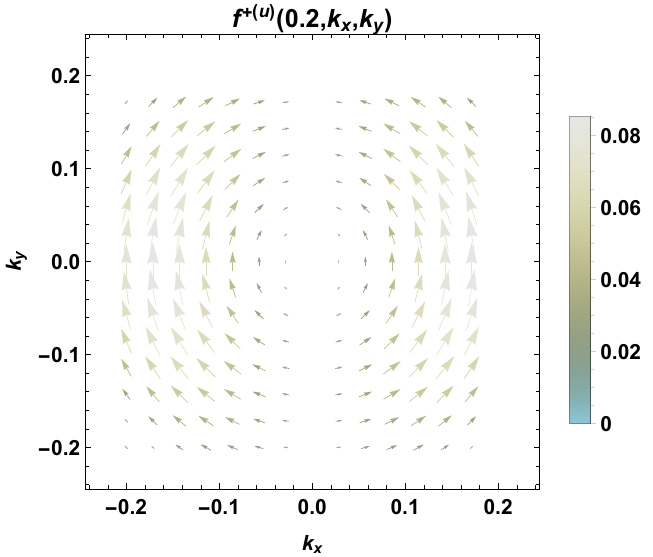}
		\hspace{0.03cm}
		(b)\includegraphics[width=5.5cm]{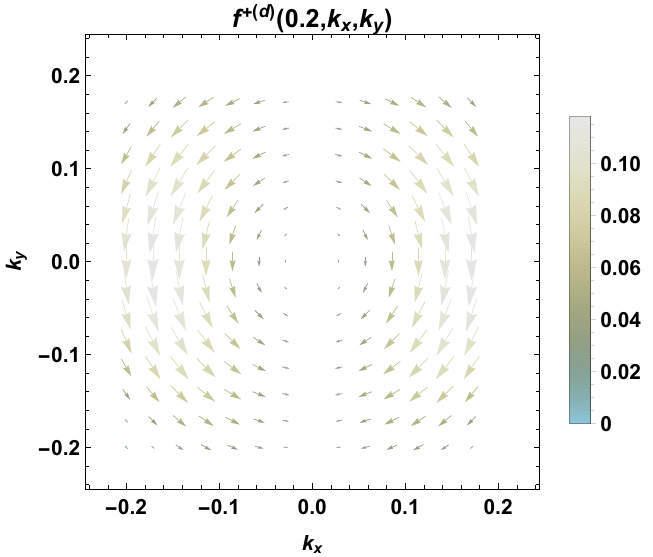}
		\hspace{0.03cm}
	\end{minipage}
	\caption{\label{figtip2_ss} 
    Third component of the TMD EMT $T^{i+}$, $f^{+ (q)}$, for (a) $u$- and (b) $d$-quark flavors of the proton in the transverse momentum $(k_x-k_y)$ plane in LFQDM.
	}
\end{figure*} 
\section{Result and Discussion \label{Section4}}
The components $\mathbb{T}^{++}_q$, $\mathbb{T}^{+i}_q$ and $\mathbb{T}^{i+}_q$ of the EMT encode important information about the valence quark structure of the proton in momentum space, and their analysis requires a detailed understanding of the leading as well as higher‑twist TMDs that enter these matrix elements. Given that the leading‑twist TMDs have already been investigated in the present frameworks in Refs. \cite{Bacchetta:2008af, Maji:2016yqo}, we begin this section by presenting the structure and behavior of the twist-3 T-even and T-odd TMDs in the adopted models.

For the case of $u$-quark flavor, T-even as well as T-odd TMDs are shown in Fig. \ref{fig3du_nk} as a function of longitudinal momentum fraction $x$ and transverse momentum $\bfk$ (GeV), computed using the LFWFs obtained in the SDM. The quantity $xf^{\perp(u)}$, presented in Fig. \ref{fig3du_nk} (a), is found to have a peak at $x=0.23$ for zero transverse momentum ($\bfk=0)$, followed by a smooth falloff as the value of transverse momentum as well as longitudinal momentum fraction increases. However, in a small $x$-domain, the distribution is found to attain a maximum at a finite value of $\bfk$ (GeV) instead of zero, and the distribution then decreases smoothly with $\bfk$ (GeV). In the LFQDM, the distribution of $x f^{\perp(u)}$ has been presented in Ref.~\cite{Sharma:2023wha}. It exhibits a peak at $x=0.31$, characterized by a sharp rise in the low-$x$ region followed by a smooth falloff. Owing to the intrinsic structure of the LFQDM, an exponential decay is observed with increasing $\bfk$ (GeV).


%
All the T-odd TMDs are plotted only for finite values of $\bfk$ (GeV) due to the presence of a factor $1/\bfk$ in their expressions, which leads to a singular behavior at $\bfk=0$ (GeV).
The T-odd distribution $x f_L^{\perp(u)}$, computed in SDM, is shown in Fig. \ref{fig3du_nk} (b) as a function of longitudinal momentum fraction $x\in[0,1]$ and transverse momentum $\bfk$ (GeV), which is restricted to the interval $\in[0.02,0.8]$.
At $\bfk=0.02$, the distribution has a peak at $x=0.24$ and falls off smoothly beyond this value of $x$. As a function of $\bfk$, this distribution monotonically falls. In Fig.~\ref{fig3du_ss} (a), $x f_L^{\perp(u)}$ is plotted within the LFQDM. The distribution is found to maintain its magnitude over the entire mid-$x$ region. The plot is primarily influenced by the dominant contribution from the scalar diquark sector, which is also responsible for the peak in the low-$\bfk$ and large-$x$ region.

In Fig. \ref{fig3du_nk} (c), the distribution of $x f_T^{\prime(u)}$, computed in SDM, is shown. The scalar and transversely polarized axial‑vector diquarks contribute with negative amplitudes. However, the positive contribution from the longitudinally polarized axial‑vector diquarks is larger in magnitude. It therefore dominates, giving rise to an overall positive distribution of $x f_T^{\prime(u)}$ in SDM. However, the TMD $x f_T^{\prime(u)}$ obtained within the LFQDM exhibits an overall negative distribution, where the dominant negative contribution from the scalar diquark sector overshadows the positive contribution from the axial-vector diquark sector, as evident in Fig.~\ref{fig3du_ss} (b). The distribution of $x f_T^{\perp(u)}$, presented in Fig.~\ref{fig3du_nk} (d), has a negative magnitude and attains its maximum when both the longitudinal momentum fraction and transverse momentum approach zero. The LFQDM, however, predicts this distribution to vanish identically.

For the case of $d$-quark flavor, T-even as well as T-odd TMDs are shown in Fig. \ref{fig3dd_nk} as a function of longitudinal momentum fraction $x$ and transverse momentum $\bfk$ (GeV), computed using the LFWFs obtained in the SDM. T-even distribution $x f^{\perp(u)}$, shown in Fig. \ref{fig3dd_nk} (a), yields a finite value at $x=0$ and $\bfk=0$ (GeV). As a function of $x$ at fixed $\bfk=0$ (GeV), it reaches a maximum at $x=0.12$, before decreasing monotonically. The distribution is prominent only near zero transverse momentum and it diminishes rapidly as $\bfk$ increases.

Unlike the finite value at the origin obtained in the SDM, the LFQDM predicts an indeterminate value of $x f^{\perp(u)}$ at $x=0$ and $\bfk=0$ (GeV). This behavior primarily arises due to the presence of $\frac{\bfk^2}{x^2}$ terms in the light-front wave functions of both the scalar and vector diquark sectors, as shown in Ref.~\cite{Sharma:2023wha}. Its $x$-dependence at fixed $\bfk=0$ (GeV) attains a peak at a comparatively larger value, $x=0.35$, relative to the SDM, before decreasing. The T-odd TMD $x f_L^{\perp(d)}$, computed within the SDM and LFQDM, is presented in Fig.~\ref{fig3dd_nk} (b) and Fig.~\ref{fig3dd_ss} (a), respectively. In the small-$x$ region, the SDM and LFQDM differ significantly in both the sign and shape of $x f_L^{\perp(d)}$. At $\bfk=0$, the SDM yields a nonzero positive value that decreases with increasing $\bfk$, whereas the LFQDM starts from zero and increases with the opposite sign.

The T-odd $x f_T^{\prime(d)}$ distribution exhibits similar qualitative behavior in both models, sharing the same sign and displaying a peak in the small-$x$, small-$\bfk$ region, followed by a smooth falloff as $x$ and $\bfk$ increase, as shown in Fig.~\ref{fig3dd_nk} (c) and Fig.~\ref{fig3dd_ss} (b). The negative distribution corresponding to the T-odd TMD $x f_T^{\perp(d)}$, computed in the SDM, is shown in Fig.~\ref{fig3dd_nk} (d). In the $x$-domain, the distribution is peaked at small values and then decreases. In the $\bfk$-domain, the distribution rises sharply as $\bfk \rightarrow 0$, owing to the presence of $1/\bfk$ in its expression. In the LFQDM, this TMD vanishes identically.

The leading-twist T-even $f_1^q$ and T-odd $f_{1T}^{\perp(q)}$ TMDs govern the EMT components $\mathbb{T}^{++}_q$ and $\mathbb{T}^{+i}_q$, which provide the longitudinal and transverse quark momentum densities. The vector density plots of $f_1^q$ and $f_{1T}^{\perp(q)}$ (including the factor from the proton's transverse polarization) are presented in Fig.~\ref{figtpp_nk} for both $u$- and $d$-quark flavors at fixed longitudinal momentum fraction $x$, computed using the LFWFs of the SDM. The density plots of $f_1^q$ for both flavors exhibit axial symmetry in transverse-momentum space, albeit with differences in spread and magnitude. As shown in Fig.~\ref{figtpp_nk} (a), the $u$-quark distribution spans the entire transverse plane, whereas Fig.~\ref{figtpp_nk} (b) shows that the $d$-quark distribution is confined to the low-transverse-momentum region. The corresponding density plots obtained in the LFQDM are presented in Fig.~\ref{figtpp_ss} (a) and Fig.~\ref{figtpp_ss} (b) for the $u$- and $d$-quark flavors, respectively. The distributions preserve axial symmetry but are more localized around the origin, differing primarily in magnitude.

The second term in Eqs.~(\ref{T++}) and (\ref{T+i}) contains the T-odd TMD $f_{1T}^{\perp(q)}$, which generates a hedgehog-like momentum-space distribution due to transverse proton polarization, as shown for the $u$- and $d$-quark flavors in Fig.~\ref{figtpp_nk} (c) and Fig.~\ref{figtpp_nk} (d), respectively. The $d$-quark distribution is more widely spread across the transverse-momentum plane with smaller magnitude compared to the more localized $u$-quark pattern. In the LFQDM, the T-odd TMD $f_{1T}^{\perp(q)}$ shows a similar trend but is more concentrated in the low-$\bfk$ region, as shown in Fig.~\ref{figtpp_ss} (c) and Fig.~\ref{figtpp_ss} (d).
 

The component $\mathbb{T}^{i+}$ of the EMT is governed solely by twist-3 TMDs. The first two terms in Eq.~(\ref{Ti+}) exhibit a density pattern similar to that of $\mathbb{T}^{+i}$. Therefore, the distributions of the remaining two TMDs are presented in Figs.~\ref{figtip1_nk}-\ref{figtip2_ss}, computed within both models.

A vortex-like structure is observed for the T-odd TMD $f_L^{\perp(q)}$, as shown in Figs.~\ref{figtip1_nk} (a) and \ref{figtip1_nk} (b) for the $u$- and $d$-quark flavors, respectively, at fixed longitudinal momentum fraction. The $u$-quark flavor exhibits a stronger transverse-momentum circulation compared to the $d$-quark flavor. In contrast, the $d$-quark contribution shows a more intricate behavior: at small transverse momentum, the vectors circulate in the same direction as the $u$-quark case, while with increasing transverse momentum the rotational flow reverses. This vortex-like structure reflects orbital-motion effects arising from the interplay between intrinsic transverse momentum and gluon exchange between the active quark and spectator. In the LFQDM, both $u$- and $d$-quark flavors also exhibit a clockwise circulation concentrated mainly in the central transverse-momentum region. However, a momentum-dependent inversion is absent in this model, as shown in Fig.~\ref{figtip1_ss}.

The fourth term in Eq.~(\ref{Ti+}) involves the T-odd TMD $f^{+(q)}$, and the corresponding density plots for the $u$- and $d$-quark flavors in the SDM are shown in Figs.~\ref{figtip2_nk} (a) and \ref{figtip2_nk} (b), respectively. The distribution exhibits a relatively weak and localized structure. The rotational symmetry is partially broken, and a dipole-like pattern emerges near the central region due to its angular dependence with respect to the transverse spin direction. The suppression of the vector magnitude away from the origin indicates that the correlation is dominated by low transverse-momentum contributions. In the LFQDM, the distribution is more widely spread in the transverse-momentum plane compared to the SDM results. Otherwise, the overall structure remains similar, with differences primarily in the vector orientation for the $u$-quark flavor.

\begin{figure*}
	\centering
	\begin{minipage}[c]{0.98\textwidth}
		(a)\includegraphics[width=5.5cm]{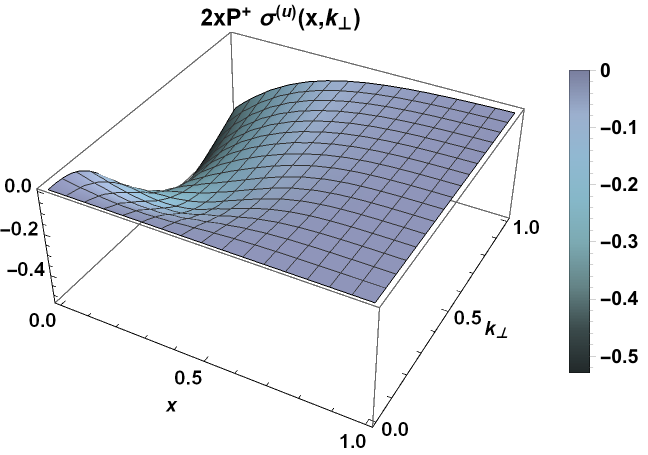}
		\hspace{0.03cm}
		(b)\includegraphics[width=5.5cm]{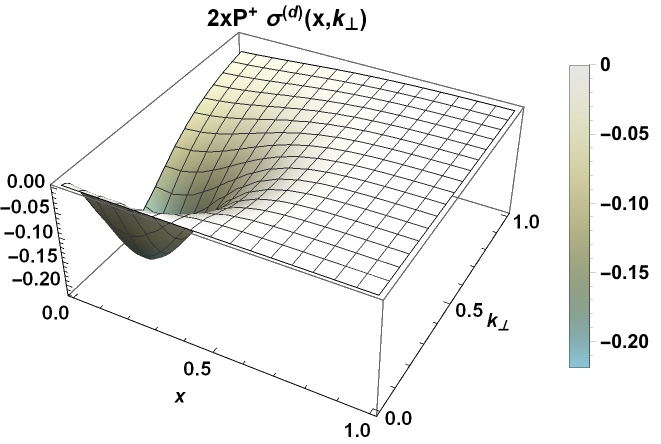}
		\hspace{0.03cm}
	\end{minipage}
	\caption{\label{3Dfigpressure_nk} 
    Distribution of transverse pressure for (a) $u$- and (b) $d$-quark flavor of proton as a function of longitudinal momentum fraction $x$ and transverse momentum $\bfk$ (GeV), computed in SDM.
	}
\end{figure*}

\begin{figure*}
	\centering
	\begin{minipage}[c]{0.98\textwidth}
		(a)\includegraphics[width=5.5cm]{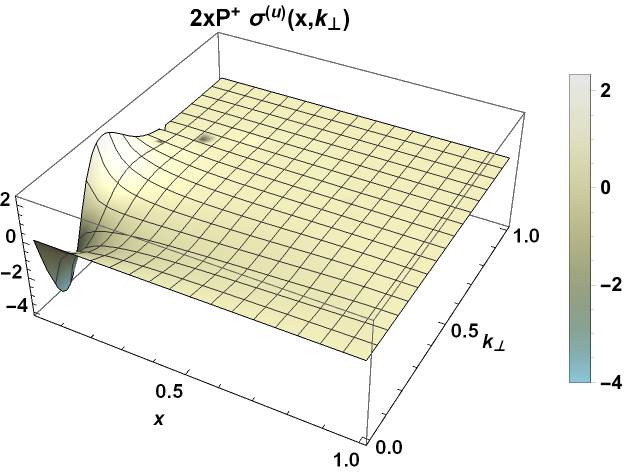}
		\hspace{0.03cm}
		(b)\includegraphics[width=5.5cm]{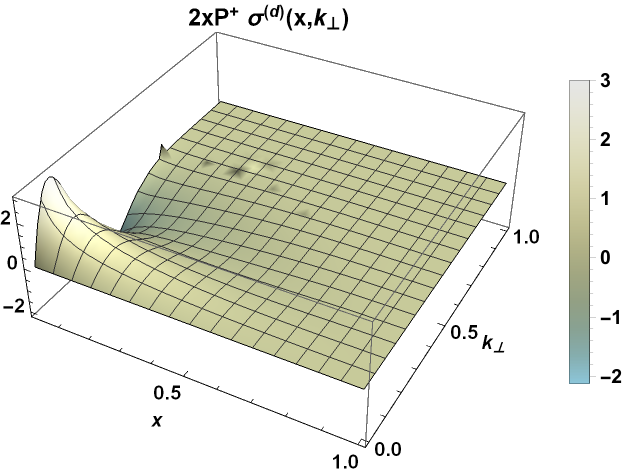}
		\hspace{0.03cm}
	\end{minipage}
	\caption{\label{3Dfigpressure_ss} 
    Distribution of transverse pressure for (a) $u$- and (b) $d$-quark flavor of proton as a function of longitudinal momentum fraction $x$ and transverse momentum $\bfk$ (GeV), computed in LFQDM.
	}
\end{figure*}

The first tensor structure associated with $\mathbb{T}_q^{ij}$ encodes information on the transverse pressure distribution of the proton in momentum space. The corresponding contributions from individual quark flavors in both the SDM and LFQDM are presented in Figs.~\ref{3Dfigpressure_nk} and \ref{3Dfigpressure_ss}, respectively, as functions of $x$ and $\bfk$ (GeV). The transverse pressure contribution from the $u$-quark flavor decreases with increasing $x$ for nonzero transverse momentum, as shown in Figs.~\ref{3Dfigpressure_nk} (a) and \ref{3Dfigpressure_ss} (a). For fixed $x$, the distribution initially increases with $\bfk$ (GeV), reaches a maximum at intermediate transverse momentum, and then decreases. The negative sign of the distribution in both models indicates confining behavior or long-range transverse correlations of the $u$-quark flavor over the full momentum range. In contrast, the positive sign obtained in the LFQDM implies a net repulsive contribution within the same kinematic region. Compared to the SDM, a more rapid falloff in both $x$ and $\bfk$ (GeV) is observed in the LFQDM for both quark flavors, and the position of the peak in $\bfk$ (GeV) is also model dependent. For the $d$-quark flavor, the SDM yields a negative distribution, as shown in Fig.~\ref{3Dfigpressure_nk}(b). In contrast, in the LFQDM the transverse pressure of the $d$-quark flavor is positive at low-$\bfk$ (GeV), but changes sign at higher transverse momentum, leading to both attractive and repulsive contributions, as shown in Fig.~\ref{3Dfigpressure_ss} (b). In summary, the SDM predicts only attractive contributions, whereas the LFQDM exhibits both attractive and repulsive behavior for both quark flavors. A similar repulsive contribution has also been reported in position space in Ref.~\cite{Won:2023zmf} at leading twist, even for both quark flavors.

\section{Conclusion \label{Section5}}
We have presented the leading and higher-twist transverse momentum-dependent parton distributions (TMDs) relevant for the study of the energy-momentum tensor (EMT) in momentum space, referred to as the TMD EMT. The corresponding momentum-space EMT correlator can be parametrized in terms of standard T-even and T-odd TMDs. To investigate the model dependence of these distributions, we have employed two light-front models based on different approaches for calculating the light-front wave functions, namely the spectator diquark model (SDM) and the light-front quark-diquark model (LFQDM). The two models differ in their momentum-space structure: the soft-wall AdS/QCD-inspired model exhibits exponential behavior, whereas the vertex-based quark-diquark model exhibits power-law momentum dependence.

In this work, the behavior of the distributions in the small-$x$ region is found to be highly sensitive to the choice of model. The sharp increase near zero transverse momentum arises from the inverse transverse momentum dependence in the twist-3 T-odd TMDs, which are computed by considering gluon exchange between the active quark and the spectator diquark. The twist-3 T-odd TMD $f_L^{\prime(u)}$ and $f_L^{\perp(d)}$ are found to have opposite signs in the two considered models, while the twist-3 T-odd TMD $f_T^{\perp(q)}$ vanishes in the LFQDM. A comparison of the vector density plots corresponding to different elements of the TMD EMT components $\mathbb{T}_q^{++}$, $\mathbb{T}_q^{+i}$, and $\mathbb{T}_q^{i+}$ is also presented. For both $\mathbb{T}_q^{++}$ and $\mathbb{T}_q^{+i}$, the element $f_1^q$ shows axial symmetry, whereas another element, $f_{1T}^{\perp(q)}$, yields a dipolar distribution. In contrast, the two elements $f_L^{\perp(q)}$ and $f^{+(q)}$ of $\mathbb{T}_q^{i+}$ exhibit vortex-like structures owing to the term $\epsilon_T^{i\textbf{k}_T}$, with one being azimuthally symmetric and the other exhibiting a dipolar deformation, respectively. The dipolar form of the distributions in the different components of the TMD EMT arises due to the choice of the transverse polarization state of the proton. Furthermore, transverse pressure contributions from both $u$- and $d$-quark flavors are investigated, which correspond to one of the tensor elements of the component $\mathbb{T}_q^{ij}$ of the TMD EMT. The SDM yields a confining contribution from both quark flavors over the entire span of $x$ and $\bfk$ (GeV). However, the LFQDM shows both attractive and repulsive behavior over the $x$-$\bfk$ range.

Hence, higher-twist parton distributions, which are usually treated as subleading contributions within the QCD factorization framework, also carry important information about the gravitational TMD structure of hadrons. At the same time, both the magnitude and sign of these distributions exhibit strong sensitivity to the underlying model, particularly in the low-$x$ and $\bfk$ (GeV) region. Even though the microscopic gravitational structure encoded in the EMT cannot be accessed directly, its connection with TMDs opens the possibility of probing certain mechanical properties of hadrons in momentum space through upcoming high-energy scattering experiments.

\section*{Acknowledgements}
H.D. would like to thank  the Science and Engineering Research Board, Anusandhan-National Research Foundation, Government of India under the scheme SERB-POWER Fellowship (Ref No. SPF/2023/000116) for financial support. 


\begin{thebibliography}{200}
	\section*{References}
	\bibitem{Collins:2004nx}
	J.~C.~Collins and A.~Metz,
	Phys. Rev. Lett. \textbf{93}, 252001 (2004).
		
	\bibitem{Polchinski:2002jw}
	J.~Polchinski and M.~J.~Strassler,
	JHEP \textbf{05}, 012 (2003).
		
		
	\bibitem{Gross:1973id}
	D.~J.~Gross and F.~Wilczek,
	Phys. Rev. Lett. \textbf{30}, 1343-1346 (1973).
		
	\bibitem{Politzer:1973fx}
	H.~D.~Politzer,
	Phys. Rev. Lett. \textbf{30}, 1346-1349 (1973).

	\bibitem{Ji:2004wu}
	X.~d.~Ji, J.~p.~Ma and F.~Yuan,
	Phys. Rev. D \textbf{71}, 034005 (2005).
    
	\bibitem{Collins:1989gx}
	J.~C.~Collins, D.~E.~Soper and G.~F.~Sterman,
	Adv. Ser. Direct. High Energy Phys. \textbf{5}, 1-91 (1989).

\bibitem{Gross:1971wn}
D.~J.~Gross and S.~B.~Treiman,
Phys. Rev. D \textbf{4}, 1059-1072 (1971).

\bibitem{Alday:2010zy}
L.~F.~Alday, B.~Eden, G.~P.~Korchemsky, J.~Maldacena and E.~Sokatchev,
JHEP \textbf{09}, 123 (2011).

	\bibitem{Qiu:1990xxa}
	J.~w.~Qiu and G.~F.~Sterman,
	Nucl. Phys. B \textbf{353}, 105-136 (1991).
		
	\bibitem{Bacchetta:2006tn}
	A.~Bacchetta, M.~Diehl, K.~Goeke, A.~Metz, P.~J.~Mulders and M.~Schlegel,
	JHEP \textbf{02}, 093 (2007).
		
	\bibitem{Meissner:2007rx}
	S.~Meissner, A.~Metz and K.~Goeke,
	Phys. Rev. D \textbf{76}, 034002 (2007).
		
	\bibitem{Meissner:2009ww}
	S.~Meissner, A.~Metz and M.~Schlegel,
	JHEP \textbf{08}, 056 (2009).
		
	\bibitem{Diehl:2015uka}
	M.~Diehl,
	Eur. Phys. J. A \textbf{52}, no.6, 149 (2016).
		
	\bibitem{Pasquini:2008ax}
	B.~Pasquini, S.~Cazzaniga and S.~Boffi,
	Phys. Rev. D \textbf{78}, 034025 (2008).

\bibitem{Burkardt:2008jw}
M.~Burkardt, C.~A.~Miller and W.~D.~Nowak,
Rept. Prog. Phys. \textbf{73}, 016201 (2010).

\bibitem{Barone:2010zz}
V.~Barone, F.~Bradamante and A.~Martin,
Prog. Part. Nucl. Phys. \textbf{65}, 267-333 (2010).

\bibitem{Aidala:2012mv}
C.~A.~Aidala, S.~D.~Bass, D.~Hasch and G.~K.~Mallot,
Rev. Mod. Phys. \textbf{85}, 655-691 (2013).

\bibitem{EuropeanMuon:1983tsy}
J.~J.~Aubert \textit{et al.} [European Muon],
Phys. Lett. B \textbf{130}, 118-122 (1983).

\bibitem{EuropeanMuon:1986ulc}
M.~Arneodo \textit{et al.} [European Muon],
Z. Phys. C \textbf{34}, 277 (1987).

\bibitem{HERMES:1999ryv}
A.~Airapetian \textit{et al.} [HERMES],
Phys. Rev. Lett. \textbf{84}, 4047-4051 (2000).

\bibitem{HERMES:2001hbj}
A.~Airapetian \textit{et al.} [HERMES],
Phys. Rev. D \textbf{64}, 097101 (2001).

\bibitem{HERMES:2002buj}
A.~Airapetian \textit{et al.} [HERMES],
Phys. Lett. B \textbf{562}, 182-192 (2003).
	\bibitem{Bastami:2020rxn}
	S.~Bastami, A.~V.~Efremov, P.~Schweitzer, O.~V.~Teryaev and P.~Zavada,
	Phys. Rev. D \textbf{103}, 014024 (2021).
		
	\bibitem{Lorce:2014hxa}
	C.~Lorc{\'e}, B.~Pasquini and P.~Schweitzer,
	JHEP \textbf{01}, 103 (2015).
	
	\bibitem{Pasquini:2018oyz}
	B.~Pasquini and S.~Rodini,
	Phys. Lett. B \textbf{788}, 414-424 (2019).
		
	\bibitem{Sharma:2022ylk}
	S.~Sharma and H.~Dahiya,
	Int. J. Mod. Phys. A \textbf{37}, 2250205 (2022).
	
	\bibitem{Sharma:2023wha}
	S.~Sharma, N.~Kumar and H.~Dahiya,
	Nucl. Phys. B \textbf{992}, 116247 (2023).

\bibitem{Boer:1997nt}
 D.~Boer and P.~J.~Mulders,
Phys. Rev. D \textbf{57}, 5780-5786 (1998).

\bibitem{Belitsky:1997zw}
A.~V.~Belitsky and D.~Mueller,
Nucl. Phys. B \textbf{503}, 279-308 (1997).

	\bibitem{Lu:2012gu}
	Z.~Lu and I.~Schmidt,
	Phys. Lett. B \textbf{712}, 451-455 (2012).
	
	
	\bibitem{Liu:2021ype}
	X.~Liu, W.~Mao, X.~Wang and B.~Q.~Ma,
	Phys. Rev. D \textbf{104}, 094043 (2021).

\bibitem{Ohnishi:2003mf}
Y.~Ohnishi and M.~Wakamatsu,
Phys. Rev. D \textbf{69}, 114002 (2004).


	\bibitem{Sharma:2024lal}
	S.~Sharma, S.~Puhan, N.~Kumar and H.~Dahiya,
	PTEP \textbf{2024}, 103B05 (2024).
    
\bibitem{Courtoy:2022kca}
A.~Courtoy, A.~S.~Miramontes, H.~Avakian, M.~Mirazita and S.~Pisano,
Phys. Rev. D \textbf{106}, 014027 (2022).

\bibitem{Pasquini:2019evu}
B.~Pasquini, S.~Rodini and A.~Bacchetta,
Phys. Rev. D \textbf{100}, 054039 (2019).

\bibitem{Lorce:2023zzg}
C.~Lorc{\'e} and Q.~T.~Song,
Phys. Lett. B \textbf{843}, 138016 (2023).

\bibitem{Bacchetta:2008af}
A.~Bacchetta, F.~Conti and M.~Radici,
Phys. Rev. D \textbf{78}, 074010 (2008).

\bibitem{Maji:2016yqo}
T.~Maji and D.~Chakrabarti,
Phys. Rev. D \textbf{94}, no.9, 094020 (2016).

\bibitem{Lorce:2012ce}
C.~Lorce,
Phys. Lett. B \textbf{719}, 185-190 (2013).

\bibitem{Lorce:2012rr}
C.~Lorce,
Phys. Rev. D \textbf{87}, 034031 (2013).

\bibitem{Chen:2008ag}
X.~S.~Chen, X.~F.~Lu, W.~M.~Sun, F.~Wang and T.~Goldman,
Phys. Rev. Lett. \textbf{100}, 232002 (2008).

\bibitem{Wakamatsu:2010cb}
M.~Wakamatsu,
Phys. Rev. D \textbf{83}, 014012 (2011).

\bibitem{Hatta:2011zs}
Y.~Hatta,
Phys. Rev. D \textbf{84}, 041701 (2011).

\bibitem{Hatta:2011ku}
Y.~Hatta,
Phys. Lett. B \textbf{708}, 186-190 (2012).

\bibitem{Lorce:2018egm}
C.~Lorc{\'e}, H.~Moutarde and A.~P.~Trawi{\'n}ski,
Eur. Phys. J. C \textbf{79}, 89 (2019).

	\bibitem{Brodsky:2002cx}
	S.~J.~Brodsky, D.~S.~Hwang and I.~Schmidt,
	Phys. Lett. B \textbf{530}, 99-107 (2002).
	
	\bibitem{Brodsky:2002rv}
	S.~J.~Brodsky, D.~S.~Hwang and I.~Schmidt,
	Nucl. Phys. B \textbf{642}, 344-356 (2002).
	
	\bibitem{Gurjar:2022rcl}
	B.~Gurjar, D.~Chakrabarti and C.~Mondal,
	Phys. Rev. D \textbf{106}, 114027 (2022).

\bibitem{deTeramond:2011aml}
G.~F.~de Teramond and S.~J.~Brodsky,
[arXiv:1203.4025 [hep-ph]].

\bibitem{Chakrabarti:2019wjx}
D.~Chakrabarti, N.~Kumar, T.~Maji and A.~Mukherjee,
Eur. Phys. J. Plus \textbf{135}, no.6, 496 (2020).





\bibitem{Won:2023zmf}
H.~Y.~Won, H.~C.~Kim and J.~Y.~Kim,
JHEP \textbf{05}, 173 (2024).
\end{thebibliography}
\end{document}